\documentclass{aa}
\usepackage[utf8]{inputenc}
\usepackage{graphicx}

\begin{document}

\title{Cloud G074.11+00.11: a stellar cluster in formation}

\author{Mika Saajasto\inst{1}
  \and Jorma Harju\inst{1}
  \and Mika Juvela\inst{1}
  \and Liu Tie\inst{2}$^{,}$\inst{3}
  \and Qizhou Zhang\inst{4}
  \and Sheng-Yuan Liu\inst{5}
  \and Naomi Hirano\inst{5}
  \and Yuefang Wu\inst{6}
  \and Kee-Tae Kim\inst{2}
  \and Ken'ichi Tatematsu\inst{7}
  \and Ke Wang\inst{8}
  \and Mark Thompson\inst{9}
  }

\institute{Department of Physics, P.O.Box 64, FI-00014, University of Helsinki, Finland
\and Korea Astronomy and Space Science Institute, 776 Daedeokdae-ro, Yuseong-gu, Daejeon 34055, Republic of Korea
\and East Asian Observatory, 660 North A'ohoku Place, Hilo, HI 96720, USA
\and Harvard-Smithsonian Center for Astrophysics, 60 Garden Street, Cambridge, MA 02138, USA
\and Academia Sinica Institute of Astronomy and Astrophysics, P.O. Box 23-141, Taipei 10617, Taiwan
\and Department of Astronomy, Peking University, Beijing 100871, People's Republic of China
\and National Astronomical Observatory of Japan, 2-21-1 Osawa, Mitaka, Tokyo 181-8588, Japan
\and Kavli Institute for Astronomy and Astrophysics, Peking University, 5 Yiheyuan Road, Haidian District, Beijing 100871, China
\and Centre for Astrophysics Research, School of Physics Astronomy $\&$ Mathematics, University of Hertfordshire, College Lane, Hatfield AL10 9AB, UK
}

\date{Received day month year / Accepted day month year}

\abstract
{We present molecular line and dust continuum observations of a Planck-detected cold cloud, G074.11+00.11. The cloud consists of a system of curved filaments and a central star-forming clump. The clump is associated with several infrared sources and $\rm H_2 O$ maser emission.}
{We aim to determine the mass distribution and gas dynamics within the clump, to investigate if the filamentary structure seen around the clump repeats itself on a smaller scale, and to estimate the fractions of mass contained in dense cores and filaments. The velocity distribution of pristine dense gas can be used to investigate the global dynamical state of the clump, the role of filamentary inflows, filament fragmentation and core accretion.}
{We use molecular line and continuum observations from single dish observatories and interferometric facilities to study the kinematics of the region.}
{The molecular line observations show that the central clump may have formed as a result of a large-scale filament collision. The central clump contains three compact cores. Assuming a distance of 2.3 kpc, based on Gaia observations and a three-dimensional extinction method of background stars, the mass of the central clump exceeds 700 $M_\odot$, which is roughly $\sim 25\%$ of the total mass of the cloud. Our virial analysis suggests that the central clump and all identified substructures are collapsing. We find no evidence for small-scale filaments associated with the cores.}
{Our observations indicate that the clump is fragmented into three cores with masses in the range [10,50] $M_\odot$ and that all three are collapsing. The presence of an $\rm H_2O$ maser emission suggests active star formation. However, the CO lines show only weak signs of outflows. We suggest that the region is young and any processes leading to star formation have just recently begun.}


\maketitle

\section{Introduction}

Our understanding of the low-mass star formation has improved considerably, but it is still unclear how high-mass stars and clusters form. High-mass stars have a high impact on their surrounding medium because of their higher luminosities, winds, and, at their later evolutionary stages, trough supernova explosions. Moreover, expanding HII regions associated
with young massive stars may trigger subsequent star formation \citep{Karr2003, Lee2007}. Observational studies of regions where high-mass stars are thought to form are challenging due to high column densities and large distances, thus requiring observations at different scales and wavelengths. On the other hand, clustered star formation can also be studied towards intermediate-mass star forming regions that are found closer to the Sun and often have lower column densities than high-mass star forming regions \citep{Alonso-Albi2009}, thus making them prime regions to study.

\textit{Herschel} observations show that filamentary clouds are important for star formation in nearby molecular clouds \citep{Andre_PPVI2014, Molinari2010}. Filaments can also have a central role in the formation of stellar clusters. As discussed by \citet{Myers2009}, clusters are often found at 'hubs' where several filaments cross \citep[see also][]{Busquet2013}. Besides accreting mass from their surroundings and giving rise to dense cores through fragmentation, anisotropic infall along filaments may help protostellar clusters to accrete their mass \citep[e.g.,][]{Yuan2018, Liu2016, Liu_H2015, Peretto2013}. 

Magnetohydrodynamic simulations show that converging flows in highly turbulent medium form filamentary structures \citep{Padoan2001}. Several filaments found around stellar clusters can have formed through this process \citep{LiuH2012, Galvan-Madrid2010}. Transient, dense filaments can also form during the collapse of a parsec-scale turbulent clump \citep{Wang2010}. In this case, gravity determines where the filaments converge, and where the stars are formed. These filaments are expected to show longitudinal velocity gradients owing to material flow towards the centre of gravity. In the latter scenario, it is not clear whether gravitationally bound, massive \citep[$M_{\rm core} > 100\,M_\odot$,][]{Giannetti2013} prestellar cores can be identified inside the collapsing clump \citep{Tan_PPVI2014}.

\begin{figure*}
\includegraphics[width=17.8cm]{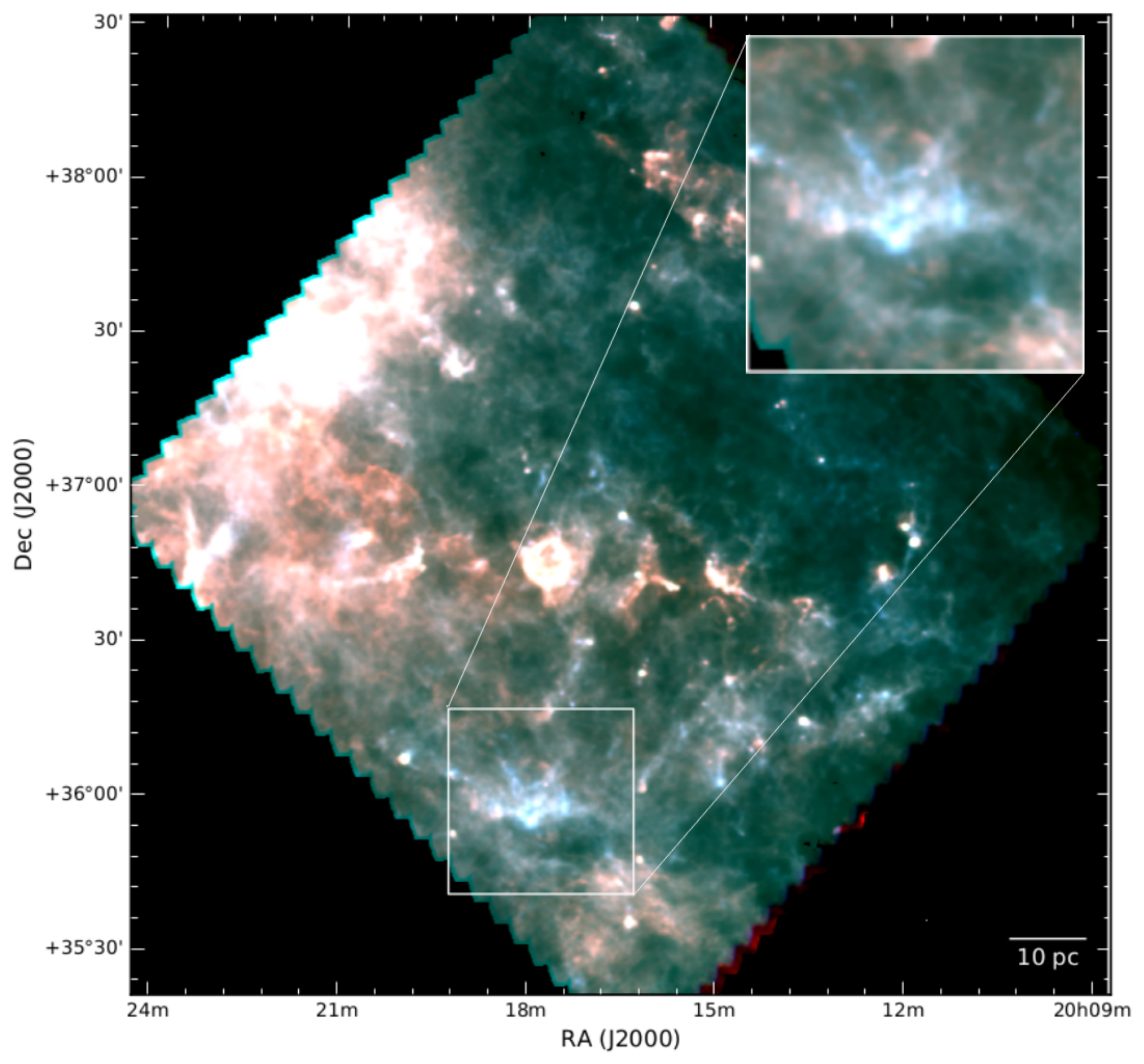}
\caption{An RGB image of the PLCKECC G074.11+00.11. The colours correspond to SPIRE and PACS observations at wavelengths 160 $\mu$m (in red), 250 $\mu$m (in green), and 350 $\mu$m (in blue). The white rectangle shows the extent of the CO observations.}
\label{fig:RGB}
\end{figure*}

We have mapped a star-forming clump residing in the Planck-detected cold cloud PLCKECC G074.11+00.11 with the Submillimetre Array (SMA), and with several single dish telescopes. The clump, associated with an intriguing system of curved filaments, shows signatures of on-going star formation, including several near infrared sources, and $\rm H_2 O$ maser emission. Following the argumentation of \citet{Liu_H2012} and \citet{Liu_H2015}, the morphology of G074.11+00.11 suggests that we are seeing a flattened, rotating cloud nearly face on. Because of the favourable orientation, relative proximity of $\sim$ 2 kpc, and relatively high mass exceeding 1000 $M_{\odot}$, G074.11+00.11 is particularly well suited for a case study, complementing statistical surveys and numerical work aimed at a better understanding of intermediate-mass and clustered star formation.


The rest of this paper is organized as follows: In Sect. \ref{Sect:2}, we give an overview of our observations. We present our main results in Sect. \ref{Sect:3} and discuss them in Sect. \ref{Sect:4}. Finally, in Sect. \ref{Sect:5} we summarise our results.

\section{Observations}\label{Sect:2}

\subsection{Dust continuum}

We used archival observations from the \textit{Herschel} Infrared Galactic Plane Survey \citep[Hi-GAL,][]{Molinari_higal}. The Hi-GAL survey used both PACS \citep{Poglitsch2010} and SPIRE \citep{Griffin2010} instruments, in parallel mode, using a scan speed of $60\arcsec$ $\rm s^{-1}$ for both instruments, to carry out a survey of the entire Galactic plane ($\pm 1^{\circ}$). The observations cover the 70, 160, 250, 350, and 500 $\mu$m bands with angular resolution of 8.5\arcsec, 13.5\arcsec, 18.2\arcsec, 24.9\arcsec, and 36.3\arcsec, respectively. An RGB image of the entire mapped region is shown in Fig. \ref{fig:RGB}. We adopted an error estimate of 10$\%$ for the \textit{Herschel} observations, which corresponds to the error estimate of PACS data when combined with SPIRE \footnote{\url{http://Herschel.esac.esa.int/twiki/bin/view/Public/PacsCalibrationWeb}}.

\begin{figure*}
\includegraphics[width=17.8cm]{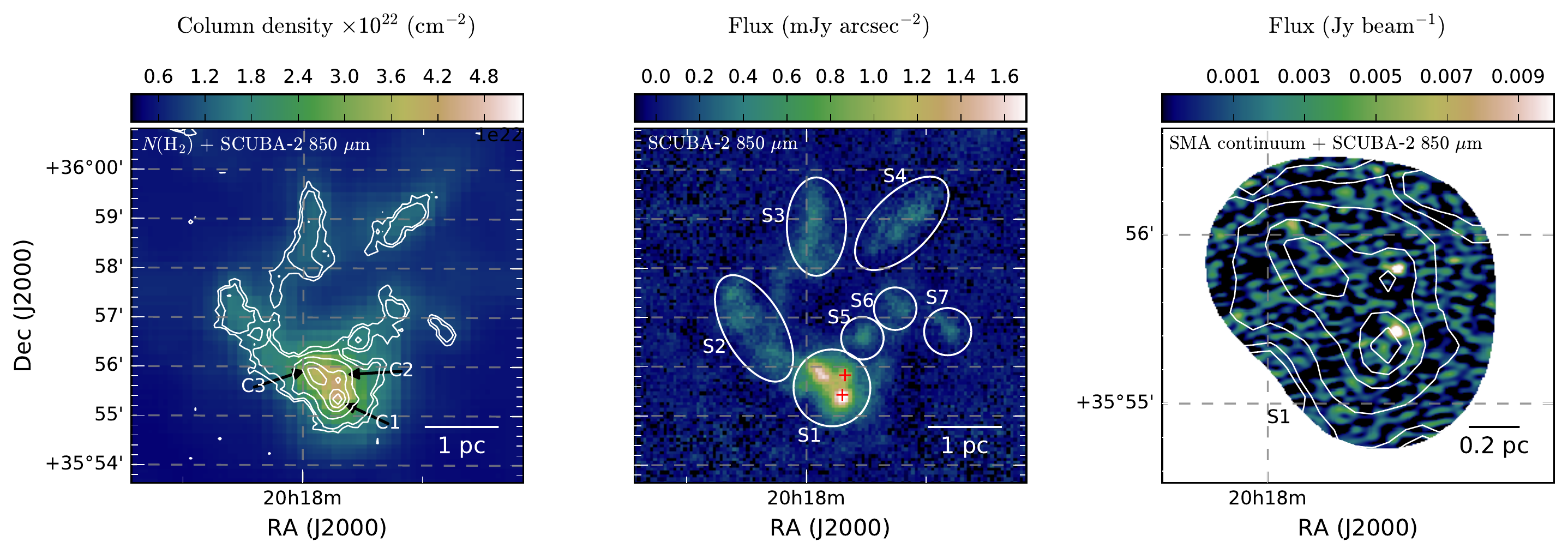}
\caption{Left panel: $\rm H_2$ column density map derived form the \textit{Herschel} observations with the SCUBA-2 observations at 850 $\mu$m shown as a contours. Labels C1-C3 indicate three compact sources revealed by SCUBA-2. Middle panel: the SCUBA-2 observations at 850 $\mu$m. The circles and ellipses, S1-S7, show the largest substructures in the field and the red crosses show the two continuum sources identified from the SMA observations. Right panel: a continuum map derived from the SMA observations, the contours show the SCUBA-2 observations at 850 $\mu$m. The area in the right panel corresponds roughly with S1.}
\label{fig:scuba}
\end{figure*}

The SCUBA-2 \citep[The Submillimetre Common User Bolometer Array 2,][]{Holland2013}, at the James Clerk Maxwell Telescope (JCMT), is a dual-wavelength camera capable of observing the same field-of-view simultaneously at 450 $\mu$m and 850 $\mu$m. G074.11+00.11 was observed by SCUBA-2 as a pilot study of the JCMT legacy survey: SCOPE  \citep[SCUBA-2 Continuum Observations of Pre-protostellar Evolution;][]{Liu2018TOPSCOPE,Eden2019}. However, the signal-to-noise level in the 450 $\mu$m band is low, only the densest peak of the clump is visible, thus we only analysed the 850 $\mu$m observations. The effective beam full width at half maximum (FWHM) of the SCUBA-2 at 850 $\mu$m is $14.1 \arcsec$ and the observations have a rms level of $\sim 0.004$ mJy $\rm arcsec^{-2}$. The 850 $\mu$m map is shown as a contours in Fig. \ref{fig:scuba}. A more detailed description of the data reduction is given by \citet{Liu2018TOPSCOPE}.

\subsection{Single dish observations}

We used observations from the Qinghai 13.7 m radio telescope of the Purple Mountain Observatory (PMO), covering the CO isotopologues in the $J = 1-0$ transition \citep{Liu2013, Meng2013, Wu2012, Liu2012}. The half power beam width (HPBW) of the telescope at 110 GHz is $\sim$ 51$\arcsec$ and the main beam efficiency is 56.2$\%$. The front end is a 9-beam Superconducting Spectroscopic Array Receiver with sideband separation \citep{Shan2012} and was used in single-sideband (SSB) mode. A Fast Fourier Transform spectrometer with a total bandwidth of 1 GHz and 16 384 channels was used as a backend. The resulting velocity resolutions of the $\rm ^{12}CO$ and $\rm ^{13}CO$ observations are 0.16 $\rm km\,s^{-1}$ and 0.17 $\rm km\,s^{-1}$, respectively. The $\rm C^{18}O$ line has a low signal-to-noise ratio and is not analysed further. The On-The-Fly (OTF) observing mode was applied, with the antenna continuously scanning a region of 22$\arcmin$ $\times$ 22$\arcmin$ with a scan speed of 20$\arcsec$ s$^{-1}$. The typical rms noise level is 0.07 K in $\rm T_{\rm mb}$ for $^{12} \rm CO$, and 0.04 K for $^{13} \rm CO$. The details of the mapping observations can be found in \citet{Liu2012} and \citet{Meng2013}.

\begin{figure}
\includegraphics[width=8.8cm]{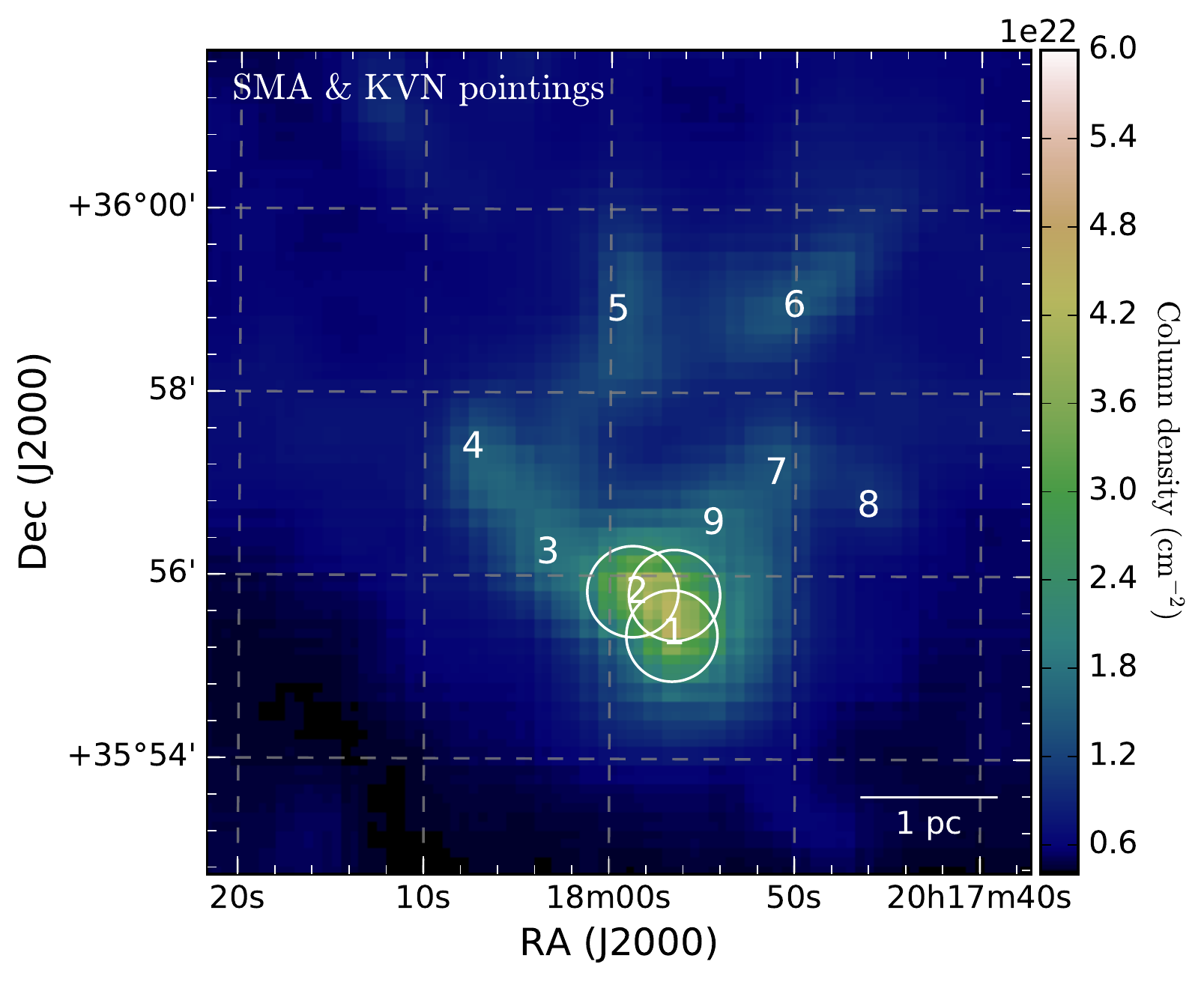}
\caption{Nine pointings observed with one of the KVN antennas, numbers from 1 to 9, and the SMA pointings shown with circles where the diameter corresponds to the FWHM of the antennas.}
\label{fig:KVN_points}
\end{figure}

Several molecular lines tracing dense gas, for example $\rm N_2H^{+}$, $\rm HCO^{+}$, were observed with a single antenna of the Korean Very Long Baseline Interferometry Network (KVN). The network consists of three 21 meter antennas with HPBW ranging from 127$\arcsec$ at 22 GHz to 23$\arcsec$ at 129 GHz. The KVN telescopes can operate at four frequency bands (i.e., 22, 44, 86, and 129 GHz) simultaneously. The main beam efficiencies are $\sim50\%$ at 22 GHz and 44 GHz and $\sim40\%$ at 86 GHz and 129 GHz The observations consist of nine pointings (see Fig. \ref{fig:KVN_points}) covering the densest region of the clump and the surrounding structures. The velocity resolution of the observations varies from 0.03 $\rm km\,s^{-1}$ to 0.21 $\rm km\,s^{-1}$ and the rms noise varies from 0.02 K to 0.04 K in $\rm T_{\rm mb}$. More details about KVN observations can be found in \citet{Liu2018TOPSCOPE}.

The $\rm ^{12}CO$ and $\rm ^{13} CO$ $J = 2-1$ transitions were observed towards the densest region of the clump with the Submillimeter Telescope (SMT) as part of the SAMPLING survey \citep[The SMT "All-sky" Mapping of PLanck Interstellar Nebulae;][]{Wang2018}. The SMT has a HPBW of $\sim 36\arcsec$ at 230 GHz and the observations have a velocity resolution of 0.33 $\rm km\,s^{-1}$ and rms noise of 0.02 K in $\rm T_{\rm mb}$ for $^{12} \rm CO$, and 0.01 K for $^{13} \rm CO$. The main beam efficiency is on average 74$\%$. All of the observed lines are summarised in Table \ref{table:lines}. 

\begin{table*}
\caption{Summary of the spectral lines that were observed with the single dish antennas.}
\centering
\begin{tabular}{c c c c c c c}
\hline\hline
& & & & & & \\
Telescope & Location & Molecule & Transition & Rest frequency & $\Delta \rm T_{\rm mb}$\tablefootmark{\rm (1)} & Detection \\
& & & & (GHz) & (K) & \\
\hline
& & & & & & \\
PMO & China & $^{12}\rm CO$ & $ J = 1 - 0 $ & 115.3 & 0.07 & Yes \\
PMO &  & $^{13}\rm CO$ & $ J = 1 - 0 $ & 110.2 & 0.04 & Yes \\
PMO &  & $\rm C^{18}O$ & $ J = 1 - 0 $ & 109.8 & 0.11 & Yes \\
SMT & USA & $^{12}\rm CO$ & $ J = 2 - 1 $ & 230.5 & 0.02 & Yes \\
SMT &  & $^{13}\rm CO$ & $ J = 2 - 1 $ & 220.4 & 0.01 & Yes \\
KVN & South Korea & $\rm H_2O$ & $ J = 6(1,6)-5(2,3)$ & \hspace{0.02in} 22.2 & 0.02 & Yes \\
KVN &  & $\rm CH_3OH$ & $ J = 7(0,7)-6(1,6)$ & \hspace{0.06in}44.1 & 0.04 & No \\
KVN &  & $\rm HCO^{+}$ & $ J = 1 - 0 $ & \hspace{0.06in}89.2 & 0.02 & Yes \\
KVN &  & $\rm N_2H^{+}$ & $ J = 1-0 $ & \hspace{0.06in}93.2 & 0.03 & Yes \\
KVN &  & $\rm H_2CO$ & $ J = 2(1,2)-1(1,1) $ & 140.8 & 0.02 & Yes \\
\hline
& & & & & \\
\end{tabular}
\tablefoot{(1) Rms values estimated over 25 channels.}
\label{table:lines}
\end{table*}

\subsection{Interferometric observations}

To determine the internal structure and velocity field inside the clump, we used the Submillimeter Array \footnote{The Submillimeter Array is a joint project between the Smithsonian Astrophysical Observatory and the Academia Sinica Institute of Astronomy and Astrophysics, and is funded by the Smithsonian Institution and the Academia Sinica.} \citep[SMA;][]{Ho_sma}. The observations cover three tracks, two in sub-compact configuration and one in compact configuration, observed in June and August 2016. The characteristic resolutions of the configurations are $\sim$ 5$\arcsec$ and 2.5$\arcsec$ at 345 GHz for the sub-compact and compact configurations, respectively. Each track consists of three pointings, covering the densest region of the clump as shown in Fig. \ref{fig:KVN_points}. In this paper we concentrate on the $\rm ^{12} CO$ and $\rm ^{13} CO$ $ J=2-1$ lines. The observations have an angular resolution of $\sim$ 3.82$\arcsec$ $\times$ 2.42$\arcsec$ (the beam position angle is $\rm bpa = 85.32^{\circ}$) and a velocity resolution of 0.17 $\rm km\,s^{-1}$ for the SWARM correlator. The sources $\rm 3C273$ and J2015$+$371 were used for bandpass and gain calibration, while Uranus and Titan were used for flux calibration. We used the Python scripts \textit{sma2casa.py} and \textit{smaImportFix.py}\footnote{\url{https://www.cfa.harvard.edu/sma/casa/}} to convert the raw SMA data to CASA format and CASA version 5.1.1 to calibrate and image the observations. Continuum visibility data at 1.3 mm were constructed by a fit to the line-free channels and the continuum was then subtracted from the spectral-line data. 

To reduce the problems caused by missing short spacing information, we used the SMT $\rm ^{12}CO$ and $\rm ^{13}CO$ $J = 2 - 1$ observation as an initial model for the CASA \textit{clean} routine. The model was created by interpolating and re-gridding the SMT observations to match the SMA observations and smoothing the velocity resolution of both SMA and SMT to 0.4 $\rm km\,s^{-1}$. The resulting \textit{clean} images were then feathered with the SMT observations to produce the final clean image cubes. The rms levels for the CO observations and the continuum are $\sim 0.038$ $\rm Jy$ $\rm beam^{-1}$ and $0.002$ $\rm Jy$ $\rm beam^{-1}$, respectively.

\section{Results}\label{Sect:3}

\subsection{Dust continuum observations}

The SPIRE observations at 250 $\mu$m, 350 $\mu$m, and 500 $\mu$m, are used to derive the dust optical depth at 250 $\mu$m and to estimate the $\rm H_2$ column density. We convolved the surface brightness maps to a common resolution of 40$\arcsec$ and applied average colour correction factors described by \citet{Sadavoy2013}. The final colour temperatures are computed by fitting the spectral energy distributions (SEDs) with modified blackbody curves with a constant opacity spectral index of $\beta = 1.8$. Although the exact value of the spectral index is know to vary from cloud to cloud, $\beta = 1.8$ can be taken as an average value in molecular clouds \citep{PlanckXXII, PlanckXXIII, PlanckXXV, Juvela2015V}. Furthermore, using a constant spectral index has been shown to reduce biases in the derived parameters \citep{Menshchikov2016}. The 250 $\mu$m optical depth is given by 

\begin{figure*}
\sidecaption
\includegraphics[width=12cm]{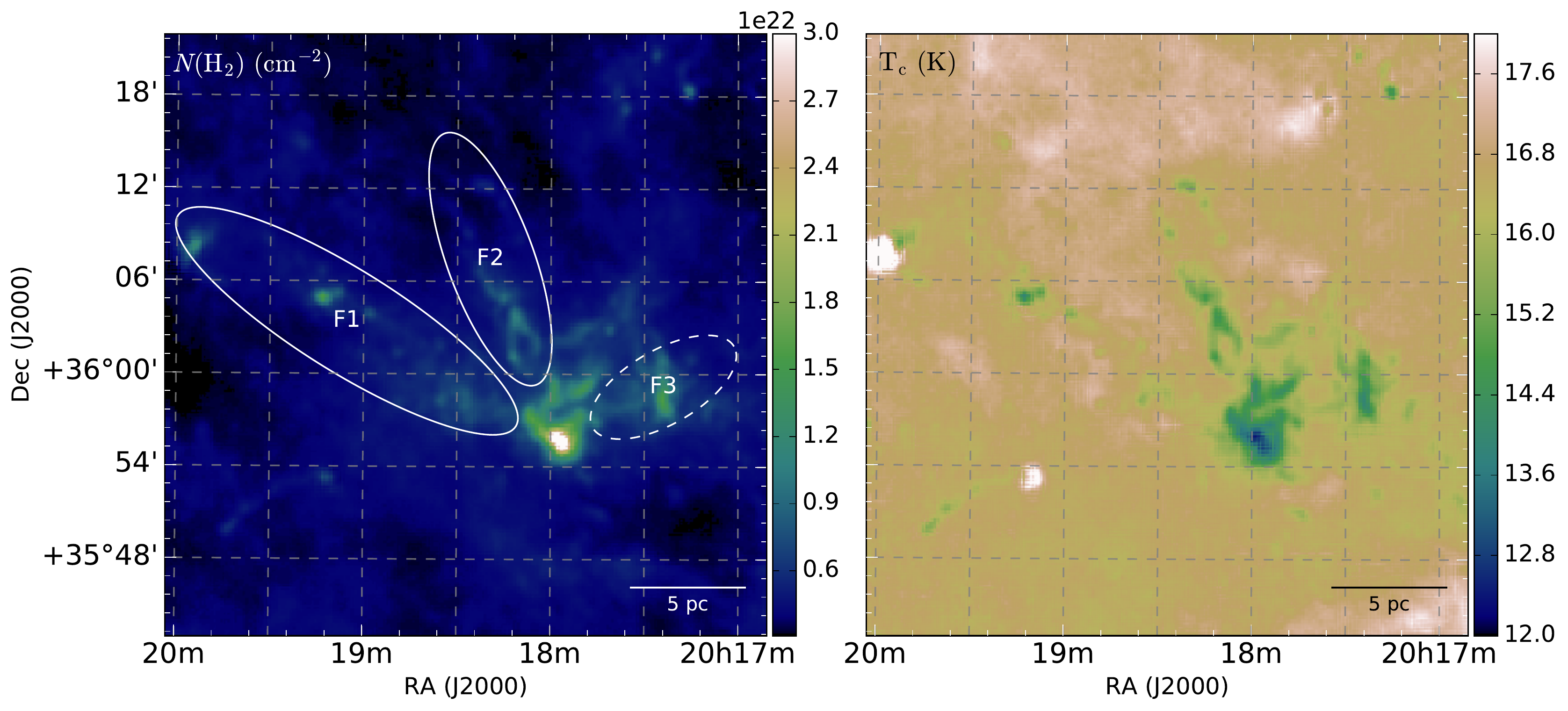}
\caption{$\rm H_2$ column density map (left panel) and a colour temperature map (right panel) derived from the \textit{Herschel} observations. The ellipses F1 and F2 show the two filamentary structures connected to the central clump. The dashed ellipse F3 shows the location of the possible third filamentary structure.}
\label{fig:nh2_temp}
\end{figure*}

\begin{equation}
\tau_{\rm 250} = \frac{I_{\nu}(\rm 250 \mu m)}{B_{\nu}(T_{\rm c})},
\end{equation}
where $I_{\nu}(\rm 250 \mu m)$ is the fitted 250 $\mu$m intensity and $T_{\rm c}$ is the colour temperature. The data were weighted according to $7 \%$ error estimates for SPIRE surface brightness measurements. The hydrogen column density $N(\rm H_2)$ was then solved from 

\begin{equation}
\tau_{\nu} = \kappa_{\nu} N(\rm H_2) \hat{\mu},
\end{equation}
where $\kappa_{\nu}$ is the dust opacity and $\hat{\mu}$ is the total gas mass per $\rm H_2$ molecule, 2.8 amu. We assume a dust opacity of $\kappa_{\nu} = 0.1(\nu / 1000$ $\rm GHz)^{\beta}$ $ \rm cm^2$ $g^{-1}$ \citep{Beckwith1990}. The resulting $\rm H_2$ column density and colour temperature maps are shown in Fig. \ref{fig:nh2_temp}. The central region of the clump reaches column density of $\sim 4.0 \times 10^{22}$ $\rm cm^{-2}$ and a temperature of $\sim12$ K. The filament-like structures around the clump are almost as dense and cold as the central region of the clump, with column densities in the range $[1-2] \times 10^{22}$ $\rm cm^{-2}$ and temperatures in the range $[14-16]$K. The assumed error estimates cause an uncertainty of $\sim 1 \%$ in the temperature and an uncertainty of $\sim 8\%$ in the column density. The filamentary structures are faint but appear to be continuous in the \textit{Herschel} images. The structures are also visible in the 12 $\mu$m map obtained by the Wide-field Infrared Explorer (WISE, see Fig. \ref{fig:WISE}), confirming that the structures are continuous and extend several parsecs away from the central clump.

The SCUBA-2 observations at 850 $\mu$m trace the densest region of the clump and clearly show smaller dense substructures (labelled S2 to S7 in Fig. \ref{fig:scuba}) above the central clump. The densest ridge of the central clump, S1, has an elliptical shape with the major axis along the northeast-southwest direction. The 850 $\mu$m emission shows three separate intensity peaks in the central region of the clump, but only one of the peaks, C1, has a corresponding point sources in the WISE maps from 3.4 $\mu$m to 12 $\mu$m. The 450 $\mu$m map suffers from poor signal-to-noise ratio (S/N), only the brightest peak of the 850 $\mu$m map can be identified, thus, we only use the 850 $\mu$m map in our analysis.

\subsection{Spectral lines}

Shown in Fig. \ref{fig:KVN_n2hp}, are the $\rm N_2H^+$ spectra from the KVN pointings shown in Fig. \ref{fig:KVN_points}. We used the GILDAS/CLASS software to make a hyperfine fit to the spectra to estimate the central velocity and the FWHM of the line. For positions 1 to 3 the central velocities increase from $-2.10 \pm 0.03$ km\,s$^{-1}$, to $-0.84 \pm 0.03$ km\,s$^{-1}$, and to $0.25 \pm 0.09$ km\,s$^{-1}$, respectively. The corresponding FWHM values are $2.42\pm 0.1$ km\,s$^{-1}$, $2.29 \pm 0.1$ km\,s$^{-1}$, and $1.5 \pm 0.3$ km\,s$^{-1}$. The spectra from positions 4-9 show variation in the central velocity, with velocities in the range [$-$2.4, 2.3] $\rm km\,s^{-1}$ and FWHM values in the range [0.7, 3.1] $\rm km\,s^{-1}$. However, the S/N of the spectra from pointings 4-9 is low, and the fit fails to converge for pointing 8, thus, the resulting hyperfine fits are not accurate. The gradual shift in velocity is also seen in the $\rm H_2 CO$ line, Fig. \ref{fig:KVN_h2co}, from $\sim -2$ $\rm km\,s^{-1}$ to $\sim 1$ $\rm km\,s^{-1}$. This can indicate a streaming motion of material flowing along the filaments towards the central clump, or away from the clump \citep{Oya2014}.

\begin{figure*}
\sidecaption
\includegraphics[width=12cm]{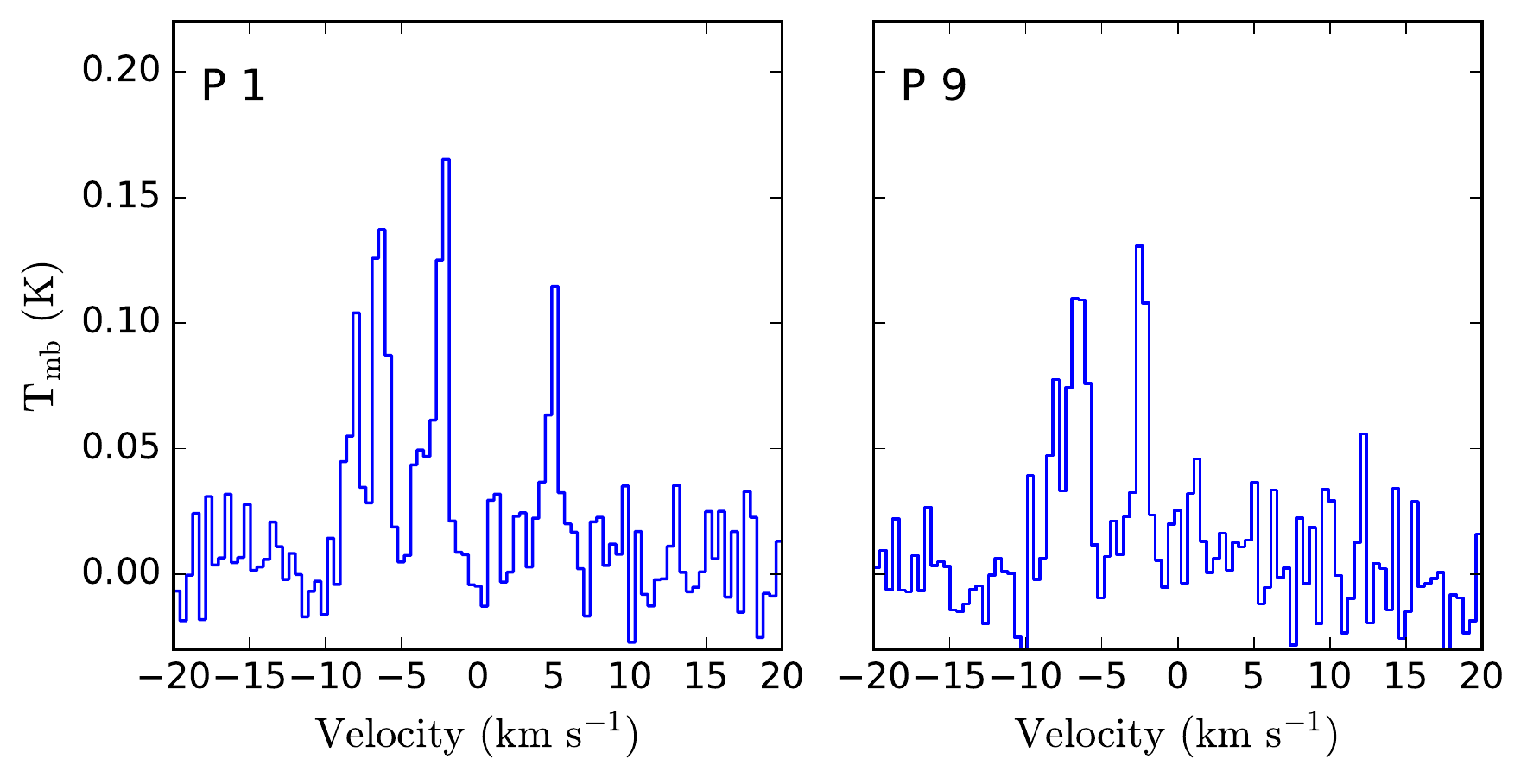}
\caption{$\rm H_2O$ maser emission detected from pointings 1 and 9 (see Fig. \ref{fig:KVN_points}) using the KVN antenna.}
\label{fig:h2o}
\end{figure*}

We detected $\rm H_2O$ maser emission at 22 GHz towards pointings 1 and 9 with the KVN antenna. The $\rm H_2O$ maser spectrum of pointing 1 (Fig. \ref{fig:h2o}), has four velocity components lying within 7 $\rm km\,s^{-1}$ from the average velocity of the dense gas, while the spectrum from pointing 9 has only two, possibly three, components, at $-$3 $\rm km\,s^{-1}$ and at $-$7 $\rm km\,s^{-1}$. We estimate the isotropic luminosity of the maser towards pointing 1, using Eq. (1) of \citet{Kim2018}:

\begin{equation}
L_{22} = 2.3 \times 10^{-8} L_\odot \frac{\int S_{22} d\nu}{\rm Jy\, km\,s^{-1}} \frac{D^2}{\rm kpc},
\end{equation}
where $S_{22}$ is the integrated flux density of the $\rm H_2O$ line at 22 GHz and $D$ is the distance to the source, for which we assumed 2.3 kpc. The distance estimate is discussed in Appendix B. The integrated flux density towards pointing 1 is $\sim 20$ $\rm Jy\, km\,s^{-1}$ and the estimated isotropic luminosity of the maser emission is $L_{22}$ $\sim 2.43 \times 10^{-6}$ $L_\odot$. Towards pointing 1, we have also a clear detection of HCO$^+$ (Fig. \ref{fig:KVN_hco}). The line profile of HCO$^+$ shows a weak broadening towards higher velocities. To quantify the asymmetry we computed an asymmetry parameter $\delta_{v}$ defined by \citet{Mardones1997}  

\begin{equation}
\delta_{v} = v_{\rm thick} - v_{\rm thin} / \Delta v_{\rm thin},
\end{equation}
where $v_{\rm thick}$ and $v_{\rm thin}$ are the central velocities of the optically thick and thin lines, for which we use the HCO$^+$ and $\rm N_2H^+$, respectively, and $\Delta v_{\rm thin}$ is the FWHM of the optically thin line. For the $\rm N_2H^+$ line we use the results of the hyperfine fit from position 1 and for the HCO$^+$ line we estimate a central velocity of $-$2.95 $\rm km\,s^{-1}$, resulting in an asymmetry value of $\delta_v = -0.35$. For asymmetric lines, \citet{Mardones1997} used the criterion $\left| \delta_v \right| > 0.25$, thus in our case, the HCO$^+$ line has a clear blue asymmetry. The detection of a $\rm H_2O$ maser and the asymmetry of the HCO$^+$ line \citep{Fuller2005}, are suggestive of star-formation activity in the clump.


\begin{figure*}
\sidecaption
\includegraphics[width=12cm]{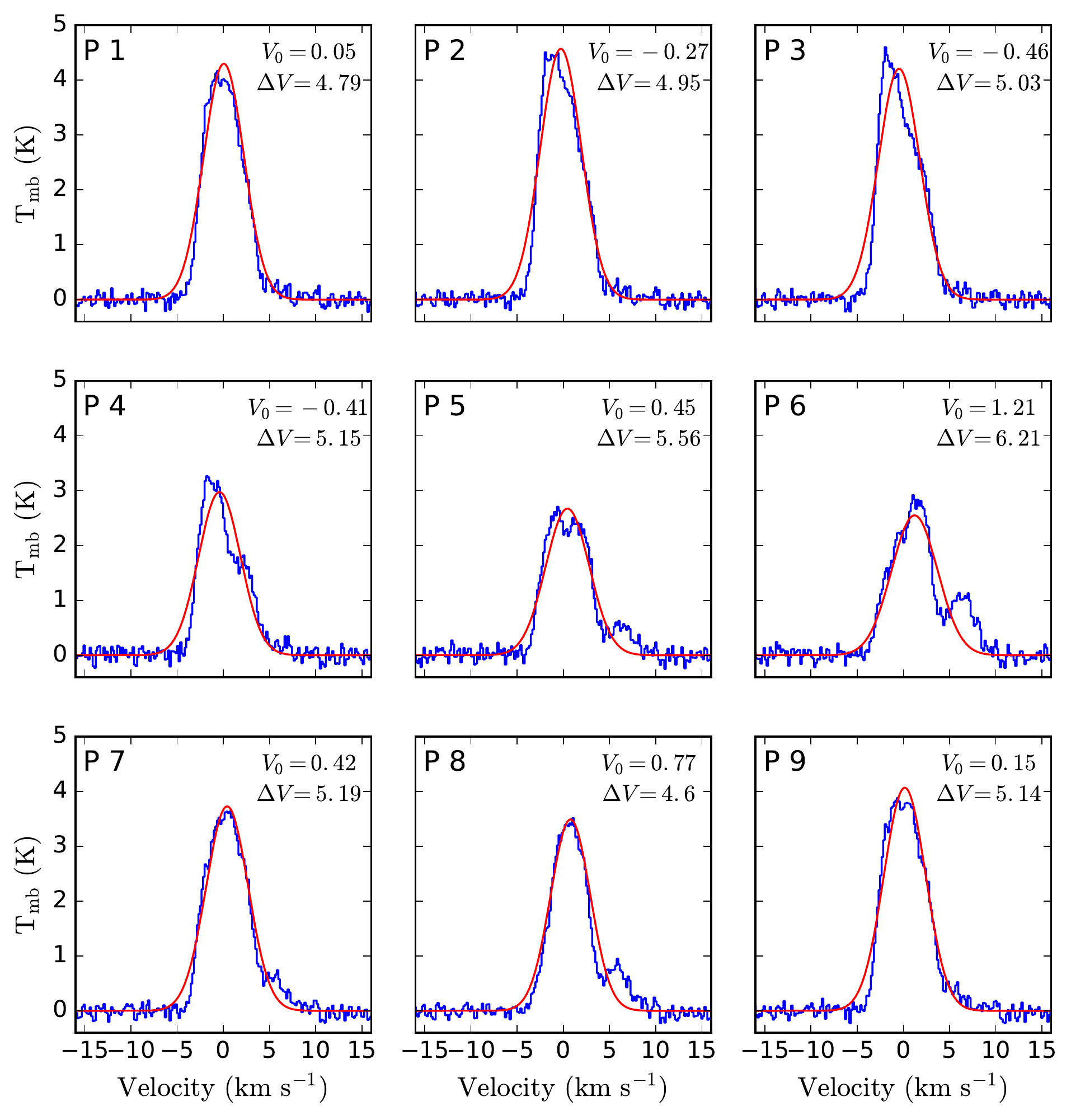}
\caption{$\rm ^{13} CO$ $J=1-0$ spectra extracted from the PMO observations towards the KVN pointings shown in Fig. \ref{fig:KVN_points}. The red line is a Gaussian fit to the line profile and the numbers in top-right corner are the Gaussian parameters of the fit in $\rm km\,s^{-1}$.}
\label{fig:Q12CO}
\end{figure*}

\begin{figure*}
\sidecaption
\includegraphics[width=12cm]{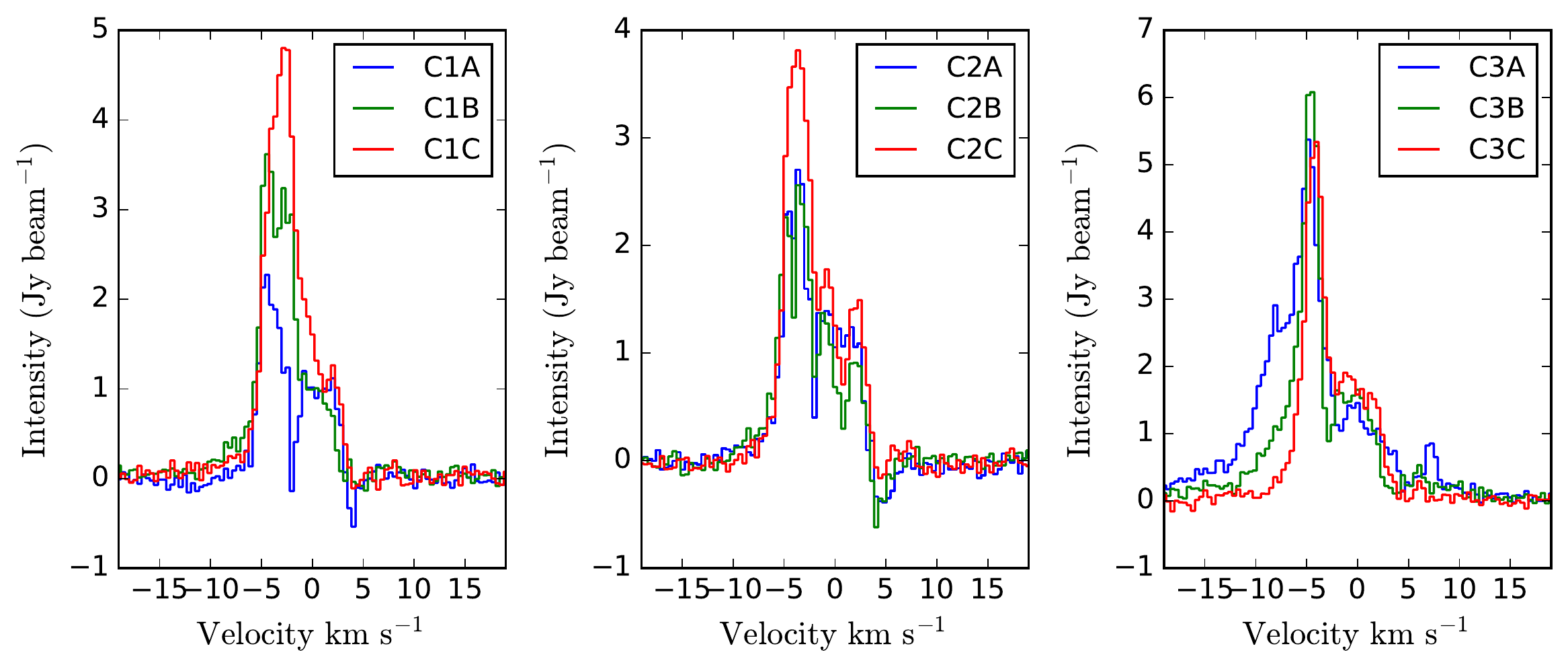}
\caption{$\rm ^{12} CO$ $J=2-1$ spectra obtained with the SMA towards positions indicated in Fig. \ref{fig:SMA_spec_locs}.}
\label{fig:SMA12CO}
\end{figure*}

\begin{figure}
\includegraphics[width=8.8cm]{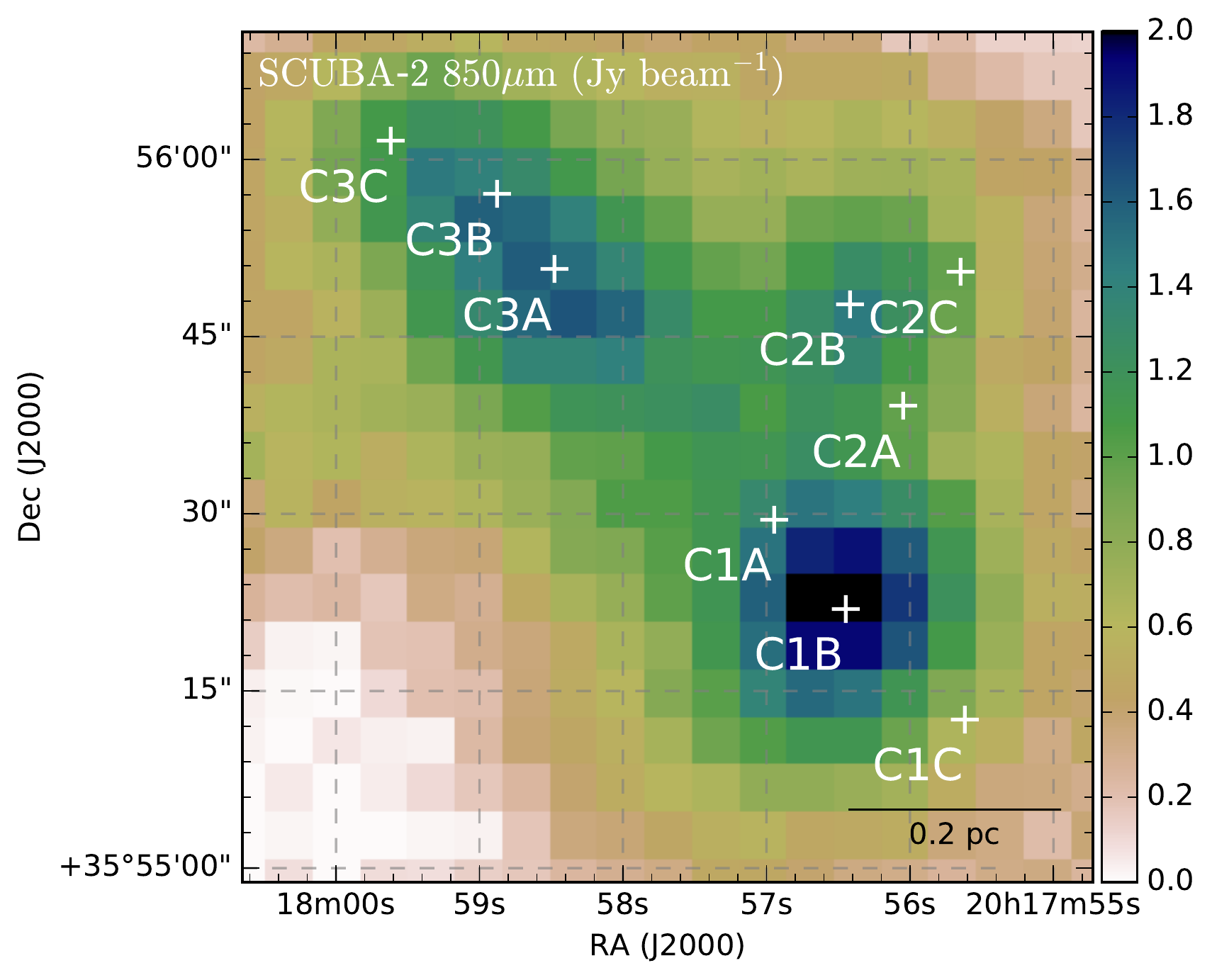}
\caption{Locations within the central clump where we have extracted the SMA $\rm ^{12} CO$ $J=2-1$ spectra (white crosses). The colour map shows the SCUBA-2 intensity at 850 $\mu$m.}
\label{fig:SMA_spec_locs}
\end{figure}

Shown in Fig. \ref{fig:Q12CO} are $^{13} \rm CO$ $J=1-0$ spectra extracted from the PMO observations towards the KVN pointings. The spectral profiles from all nine points are very broad showing emission in the range [$-$5,5] $\rm km\,s^{-1}$. Several line profiles show clear asymmetry and can not be fitted with a simple Gaussian profile. Furthermore, the spectra from the northern side of the region, P5-P8, show a separate velocity component at $\sim 7$ $\rm km\,s^{-1}$. Spectra from SMA $^{12} \rm CO$ observations (Fig. \ref{fig:SMA12CO}) have been extracted from the locations shown in Fig. \ref{fig:SMA_spec_locs} which show similar broad emission as the PMO observations. The main emission in the SMA observations is in the range [$-$5,0] $\rm km\,s^{-1}$ and the spectral profiles show clear inverse P-Cygni profiles for cores C1  and C2, an indication of infalling material. Furthermore, the SMA spectra C3A and C3B have broad wings extending up to $\sim$ $-$15 $\rm km\,s^{-1}$, which can indicate outflows from the compact source C3. The spectra from C1 and C2 also have wings in the line profile, but they are not as broad as for C3. The spectra C3A also has a separate weak velocity component at 7 $\rm km\,s^{-1}$. 

\subsection{Kinematics}

\begin{figure*}
\includegraphics[width=17.8cm]{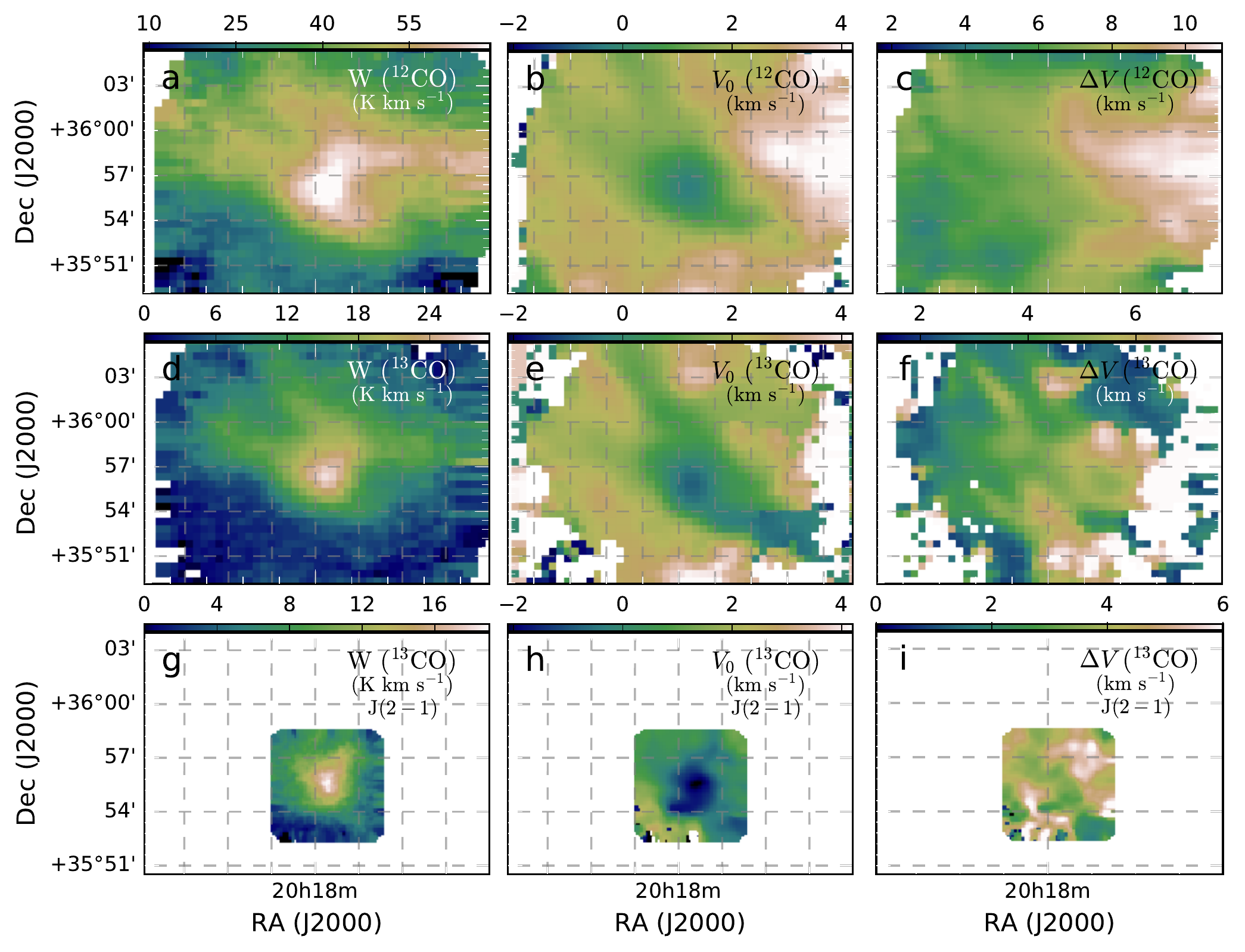}
\caption{Integrated intensity maps, the average velocity, and line width maps derived from Gaussian fits. The maps are derived form the PMO $\rm ^{12}CO$ and $\rm ^{13}CO$ $J=1-0$ and SMT $\rm ^{13}CO$ $J=2-1$ observations, panels A-F and G-I, respectively.}
\label{fig:moments}
\end{figure*}

Integrated intensity and velocity maps from Gaussian fits to the PMO $\rm ^{12}CO$ and $\rm ^{13}CO$ $J=1-0$ observations show two or three filamentary structures and the central clump at clearly lower average velocity (Fig. \ref{fig:moments} A-F). The integrated intensity maps are computed over the range [$-6,10$] $\rm km\,s^{-1}$ (panels A and D), and show that the central clump appears elongated along south east-north west direction, while two of the filamentary structures point towards north-east(F1 and F2 in Fig. \ref{fig:nh2_temp}). A possible third filament F3 extends towards west. However, it is not evident in the $^{13} \rm CO$ velocity maps, furthermore the line width maps of the $\rm ^{12} CO$ line show extremely broad line widths and a higher velocity at the western side of the region compared to the clump or filaments F1 and F2. Thus the filament F3 is likely a completely separate velocity component on the same line of sight. 

Shown in Fig. \ref{fig:moments} G-H are the moment maps of the $\rm ^{13}CO$ $J=2-1$ observations from the SMT. Compared to the $\rm ^{13}CO$ observations from PMO, the integrated intensity and the velocity maps are similar showing wider line widths on the eastern side of the clump. However, there appears to be a coherent structure at the southern side of the clump in the velocity and the line width maps, which is not seen in the PMO maps. Furthermore, in the integrated intensity map (integrated over the range [$-5.5,4.5$] $\rm km\,s^{-1}$), panel G, the clump appears to be elongated towards north west-south east, whereas in the dust continuum maps (Fig. \ref{fig:scuba}), the clump is more elongated in the north east-south west direction. The FWHM maps from both SMT and PMO, Fig. \ref{fig:moments} panels C, F, and I, show coherent line widths for the clump and the filaments F1 and F2.

\begin{figure*}
\sidecaption
\includegraphics[width=12cm]{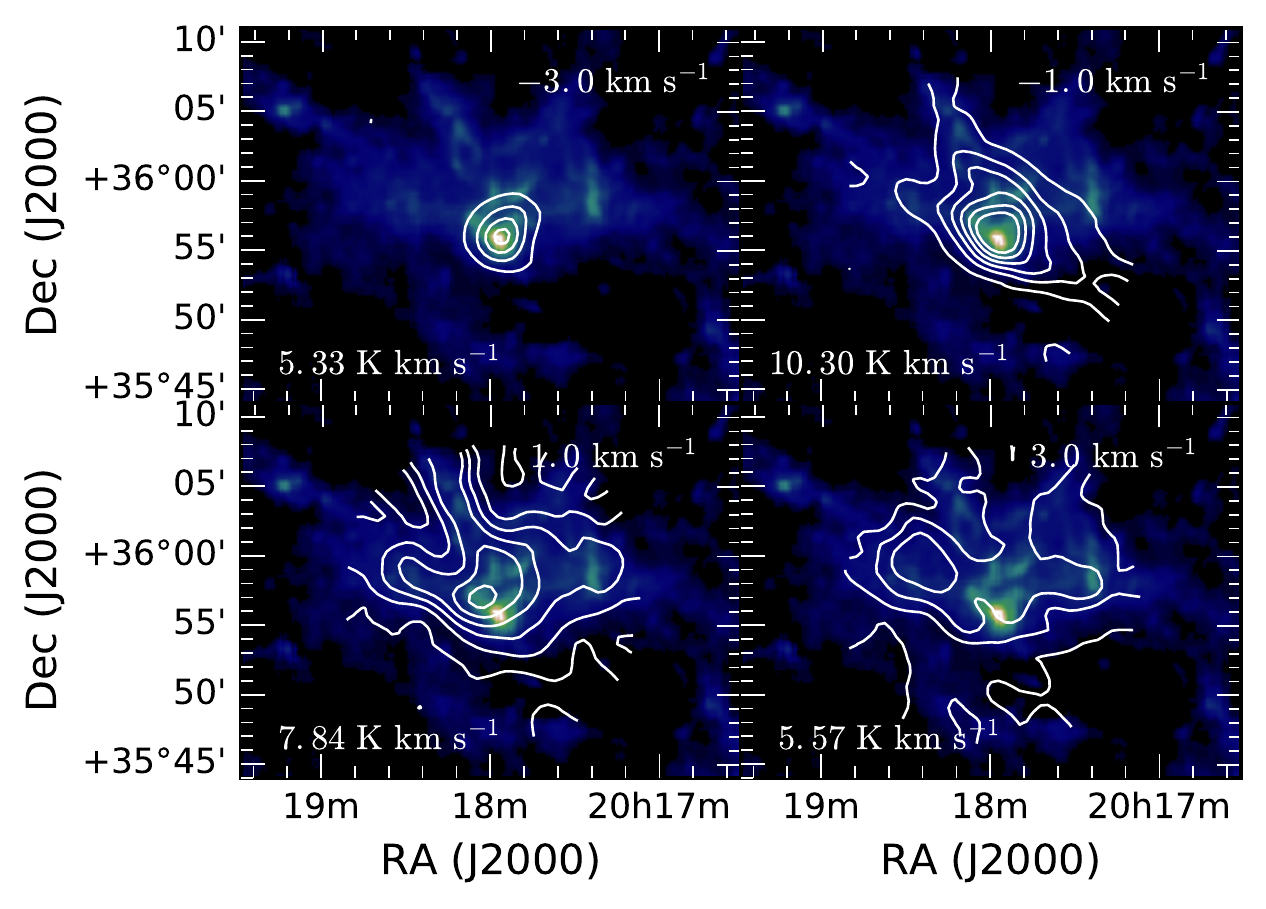}
\caption{Channel maps of the $^{13}\rm CO$ $J=1-0$ line emission as observed with the PMO (white contours), superposed on the $N(\rm H_2)$ map. The line intensities are integrated over 2.0 km\,s$^{-1}$ wide velocity intervals. The velocity centroids are indicated in the top right of each panel and the numbers in the bottom left show the maximum integrated intensity of the bin.}
\label{fig:Q13CO}
\end{figure*}

Shown in Fig. \ref{fig:Q13CO} are the PMO $\rm ^{13}CO$ $J = 1-0$ line observations integrated over 2 $\rm km\,s^{-1}$ wide velocity intervals. The $\rm ^{13}CO$ line shows a large spread in velocity, ranging from $-$4 $\rm km\,s^{-1}$ to 5 $\rm km\,s^{-1}$, with the main emission seen in range $[-2,0]$ $\rm km\,s^{-1}$ and the central clump is clearly seen in the range $[-4,0]$ $\rm km\,s^{-1}$. Overall the $\rm ^{13}CO$ line is concentrated around the main clump and traces the two or three low density filaments. The $\rm ^{13}CO$ $J = 2-1$ channel maps, Fig. \ref{fig:PMO_13CO}, are similar to Fig. \ref{fig:Q13CO}, showing the clump in the range $[-4,0]$ $\rm km\,s^{-1}$, and tracing the filamentary structure F2 in the range [$0-2$] $\rm km\,s^{-1}$.

\begin{figure*}
\sidecaption
\includegraphics[width=12cm]{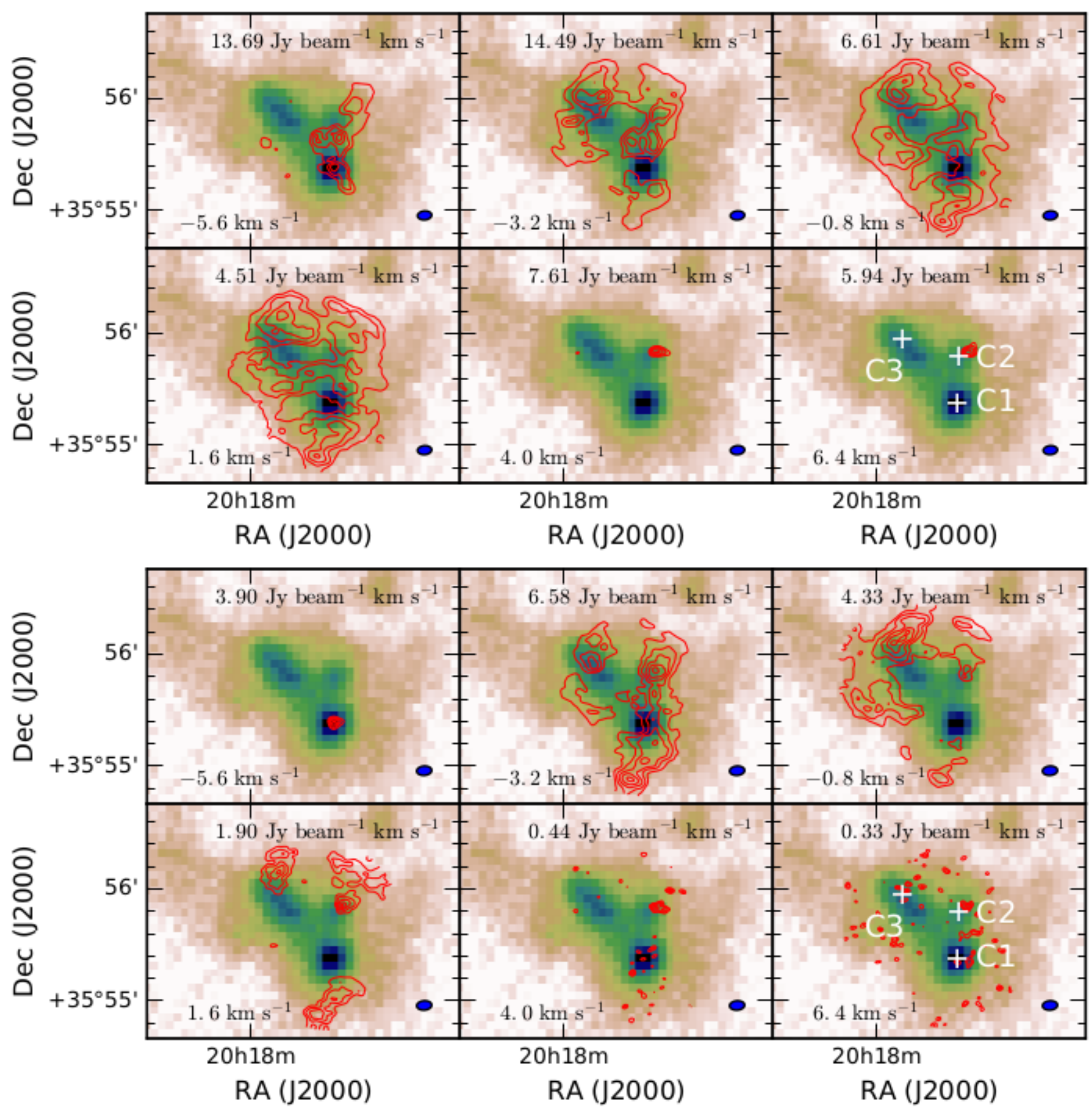}
\caption{SMA $^{12}\rm CO$ (upper row) and $^{13}\rm CO$ (lower row) $J=2-1$ line observations (red contours) integrated over  2.4 km\,s$^{-1}$ wide velocity intervals and plotted over the SCUBA-2 observations at 850 $\mu$m. The numbers in the lower left corner show the center of the velocity bin and the numbers at the top show the maximum intensity of the bin. The beam size is indicated by the blue ellipse in the lower right corner. The white crosses show the location of the three compact sources.}
\label{fig:SMA_conts}
\end{figure*}

Shown in Fig. \ref{fig:SMA_conts} are the SMA $\rm ^{12}CO$ and $\rm ^{13}CO$ $J=2-1$ line observations integrated over 2.4 km\,s$^{-1}$ wide velocity intervals. The $\rm ^{13} CO$ channel maps reveal three compact sources in the range $[-5,1.5]$ $\rm km\,s^{-1}$, all associated with an emission peak in the SCUBA-2 map. The contour maps do not show any steep velocity gradients and they do not show any evidence of filamentary structures in the largest velocity bins as seen in the PMO observations, indicating a more isotropic accretion on the point sources from the surrounding medium. The SMA $\rm ^{12}CO$ line shows more of the diffuse emission but, similar to the $\rm ^{13}CO$ observations, show no trace of the filamentary structures continuing within the clump. 

\begin{figure*}
\includegraphics[width=17cm]{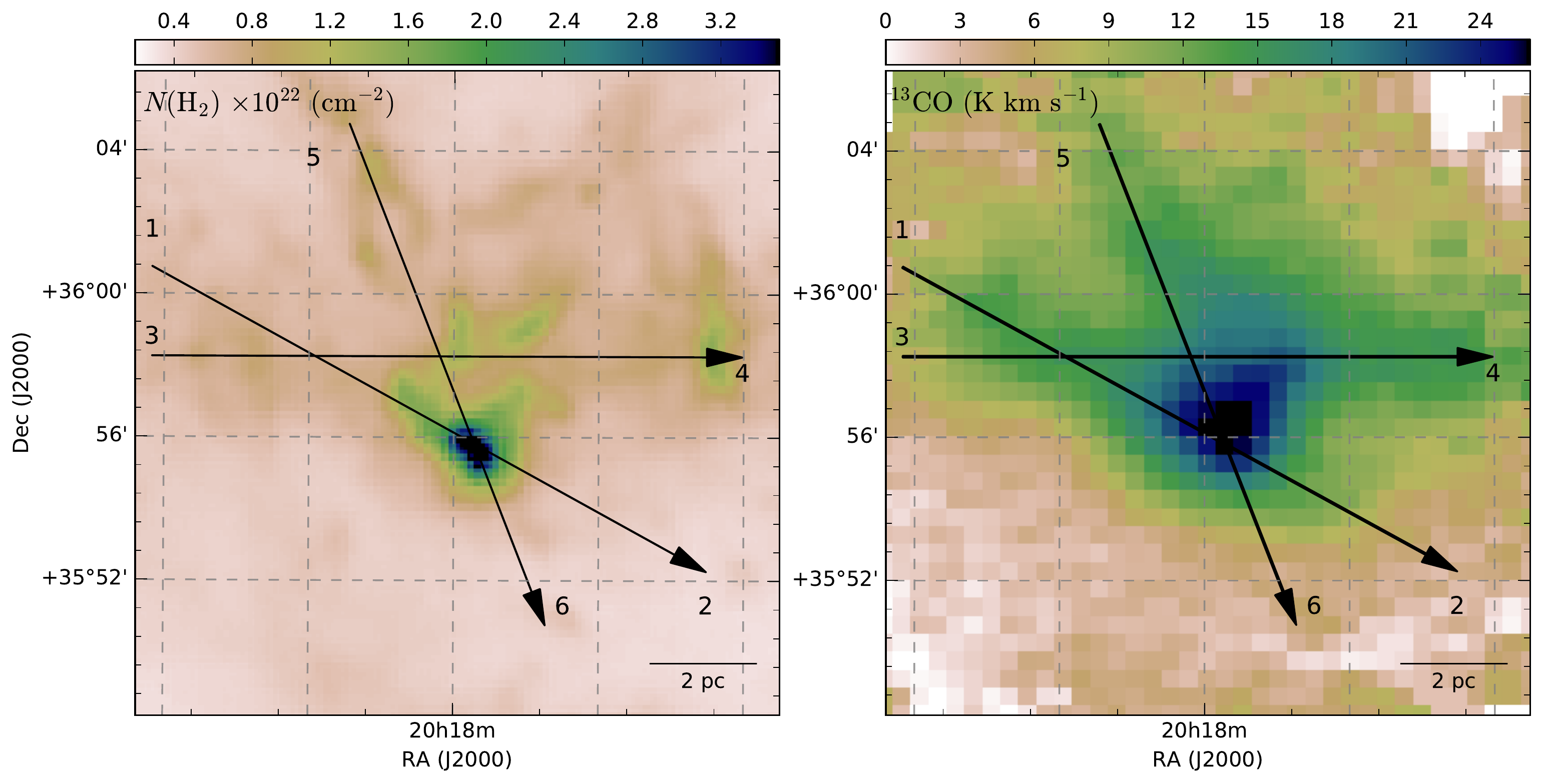}
\caption{Positions and directions (black arrows) of the position-velocity maps shown in Fig. \ref{fig:PV_13}.}
\label{fig:PV_locs}
\end{figure*}

\begin{figure*}
\includegraphics[width=17cm]{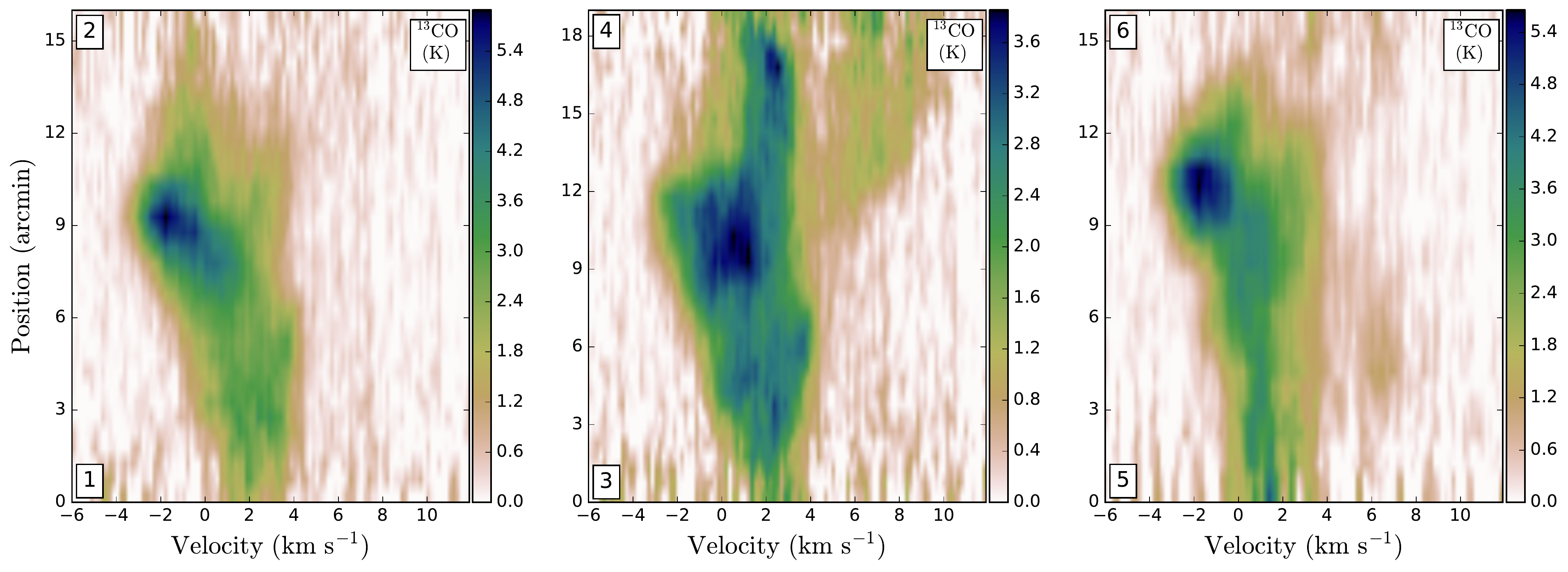}
\caption{Position-velocity maps of the $^{13}\rm CO$ $J=1-0$ line drawn along the lines shown in Fig. \ref{fig:PV_locs}.}
\label{fig:PV_13}
\end{figure*}

To study possible velocity gradients in the field we draw position-velocity (PV) maps along the lines shown in Fig. \ref{fig:PV_locs}. The cuts were chosen to trace structures that appear continuous and filamentary in the $N(\rm H_2)$ map or in the total integrated intensity map of the $\rm ^{13} CO$ line. For the PV maps in Fig. \ref{fig:PV_13}, we use the PMO $\rm ^{13}CO$ $J=1-0$ observations. In all three cuts a filamentary structure can be seen in the range $[0-2]$ km\,s$^{-1}$ and the emission is seen as a continuous structure. The central clump is at a lower velocity $\sim [-2,-1]$ $\rm km\,s^{-1}$ than the filamentary structures. Furthermore, in the vicinity of the clump, the $\rm ^{13}CO$ emission is clearly curved towards the clump, a possible indication of infall or of material flowing out from the clump \citep{Lee2000}. The cut from point 3 to point 4 clearly shows a separate velocity component at higher velocity, in the range [6$-$8] km\,s$^{-1}$, as previously shown.



\subsection{Mass estimates}

We compute the mass of the clump from both the \textit{Herschel} observations, and the SCUBA-2 observations. The mass can be derived from the total hydrogen column density following the equation

\begin{equation}\label{eq:mass_clump}
M = N(\rm{H_2}) \textit{A}  \hat{\mu} ,
\end{equation}
where $\hat{\mu}$ is the total gas mass per $\rm H_2$ molecule, 2.8 amu and $A$ is the area of the region. With the assumed distance, the mass of the central clump above a threshold of $N(\rm H_2) = 2 \times 10^{22}$ cm$^{-2}$ is $\sim 700 M_\odot$. To estimate the virial state of the region, we use the $\rm N_2H^+$ line and the equation

\begin{equation}
M_{\rm virial} = \frac{k \sigma^2 R}{\rm G}
\end{equation}
\citep{MacLaren}, where $k$ depends on the assumed density distribution, G is the gravitational constant, and $R$ is the radius of the region. We assumed a density distribution with $\rho(\rm r) \propto \rm r^{-1.5}$, and thus, $k \approx 1.33$ \citep{Chandler2000,Whitworth2001}. The total velocity dispersion, $\sigma$, including thermal and non-thermal components, was calculated from the following equation \citep{Shinnaga2004}

\begin{equation}\label{eq:sigma}
\sigma = \sqrt{\frac{k_b T_{\rm kin}}{\mu} + \frac{\Delta V^2}{8 \ln 2} - {\frac{k_b T_{\rm kin}}{m}}},
\end{equation}
where $\Delta V$ is the FWHM of the line used, $m$ is the mass of the observed molecule, and $\mu$ is the mean molecular mass, 2.33 amu, assuming $10\%$ helium. We assumed a constant kinetic temperature of $T_{\rm kin} = 15$ K, which is an average value over the region derived from the \textit{Herschel} observations. The total hydrogen column density, the shapes and radii of the regions, masses and virial masses of the regions are summarised in Table \ref{table:clump_masses}. We use a Monte Carlo method to compute the error estimates and assume  normally distributed errors. The virial mass estimates for regions 4-7 are uncertain due to low S/N ratio of the $\rm N_2H^+$ observations.

\begin{table*}
\caption{Physical parameters of the regions indicated in Fig. \ref{fig:scuba}. The shape of the region, radius, average $\rm H_2$ column density, velocity dispersion, estimated clump mass, and virial mass. A distance of 2.3 kpc pc is assumed in the mass estimates of columns 6 and 7.}
\label{table:clump_masses}
\centering
\begin{tabular}{c c c c c c c}
\hline\hline
& & & & &  &\\
& & Radius & $N(\rm H_2)$\tablefootmark{\rm (1)} & $\sigma_{\rm N_2H^+}$ & $M_c$\tablefootmark{\rm (1)} & $M_{\rm virial}(\rm N_2H^+)$ \\
Region & Shape & $\arcsec$ & $ (10^{22}$ $\rm cm^{-2})$ & (km $\rm s^{-1}$) & $(M_{\odot})$ & $(M_{\odot})$ \\
\hline
& & & & & &\\

S1 & Circle  & 54 			 & 2.63 $\pm$ 0.03 & 1.02 $\pm$ 0.22 & 700 $\pm$ 19 & 220 $\pm$ 13 \\
S2 & Ellipse & 72 $\times$ 36 & 1.44 $\pm$ 0.02 & 0.67 $\pm$ 0.31 & 340 $\pm$ 11 &  10 $\pm$  2 \\
S3 & Ellipse & 36 $\times$ 60 & 1.13 $\pm$ 0.03 & 0.37 $\pm$ 0.15 & 220 $\pm$  7 &   3 $\pm$  1 \\
S4 & Ellipse & 72 $\times$ 36 & 1.12 $\pm$ 0.01 & 0.82 $\pm$ 0.30 & 260 $\pm$ 11 &  14 $\pm$  3 \\
S5 & Circle  & 35 			 & 1.57 $\pm$ 0.02 & 1.34 $\pm$ 0.41 &  80 $\pm$  8 &  37 $\pm$  8 \\
S6 & Circle  & 35 			 & 1.26 $\pm$ 0.04 & 0.66 $\pm$ 0.19 &  65 $\pm$  6 &   8 $\pm$  2 \\
S7 & Circle  & 40 			 & 0.93 $\pm$ 0.05 & 0.71 $\pm$ 0.11 &  50 $\pm$  2 &  10 $\pm$  3 \\

\hline
& & & & & &\\
\end{tabular}
\tablefoot{(1) Estimated from the \textit{Herschel} observations.}
\end{table*}

The mass of the region can also be derived from the SCUBA-2 observations. Following \citet{Hildebrand1983}, the mass of the region is proportional to the integrated flux density

\begin{equation}
M_{850} = \frac{S_d D^2}{\kappa_{850}B_{850}(T_d)},
\end{equation}
where $S_d$ is the flux density, $D$ is the distance to the cloud, $B_{850} (T_d)$ is the Planck function at 850 $\mu$m with dust temperature $T_d$, and $\kappa_{850}$ is the opacity at 850 $\mu$m. We use a constant dust temperature of $T_d$ = 15 K and for dust opacity we assume $\kappa_{\nu} = 0.1(\nu / 1000$ $\rm GHz)^{\beta}$ $ \rm cm^2$ $g^{-1}$ \citep{Beckwith1990}.


Integrating the surface brightness of the 850 $\mu$m band over the central clump and assuming a distance of 2.3 kpc gives an estimated mass of $M_{850} =$ 230 $M_\odot$. The discrepancy between the SCUBA-2 estimate and the estimate of $\sim700$ $M_\odot$ derived from the \textit{Herschel} observations, is caused by the spatial filtering of the SCUBA-2 maps \citep{Chapin2013}. The level of filtering depends on the source properties, with extended source suffering from heavier filtering \citep{Ward-Thompson2016,Liu2018TOPSCOPE}. The SCUBA-2 850 $\mu$m map is also used to compute the masses of the three compact sources (C1, C2, and C3 in Fig. \ref{fig:scuba}). Assuming a simple circular geometry for the sources, the resulting masses are 36, 13, and 49 $M_\odot$, for C1, C2, and C3, respectively. The estimated virial masses of the compact sources are 2.0, 1.6, and 2.95 $M_\odot$, where we have used the SMA $\rm ^{13}CO$ $J=2-1$ observations and Eq. \ref{eq:sigma} to compute the velocity dispersion.

\subsection{Young stellar object}

The central clump is associated with the IRAS source 20160+3546, which is well aligned with the compact source C1. We estimate the bolometric luminosity of the IRAS source by making a modified blackbody fit to the dust continuum observations from both \textit{Herschel} and SCUBA-2, covering wavelengths from 70 $\mu$m to 850 $\mu$m. The intensity at each wavelength is computed from a circular aperture with a radius corresponding to two times the FWHM of the observations and assuming a constant opacity spectral index of $\beta = 1.8$. The resulting best-fit temperature is $T_f \sim 25$K and the luminosity of the IRAS source is $\sim 15$ $L_\odot$.

\section{Discussion}\label{Sect:4}

\subsection{Cloud structure}
Based on the \textit{Herschel} and WISE images, the clump is associated with parsec-scale, curved filamentary structures or arms that appear to be crossing at the location of the clump. The clump in turn may have formed as a result of a collision of those large scale structures, as discussed in Sect. \ref{sec:fila}. The clump has further fragmented to three cores. Assuming a distance of 2.3 kpc (see Appendix B), the total mass of the clump is  $\sim$700 $M_{\odot}$, which is $\sim$ 30$\%$ of the total mass of the cloud, while the curved filamentary structures have about 1/3 of the mass of the clump. The SCUBA-2 observations have revealed three compact sources within the clump with estimated masses in range [10,50] $M_\odot$.

\subsection{Filamentary structures}\label{sec:fila}
The low resolution $^{12} \rm CO$ and $^{13} \rm CO$ $J=1-0$ observations show that the central clump is located at a crossing of two, possibly three, large-scale filaments. The line emission shows a large spread in velocities over a range of $\sim$ 10 $\rm km\,s^{-1}$ over the whole region, while the main emission around the clump is seen in range $[-4,0]$ $\rm km\,s^{-1}$. The two main filamentary structures are seen in the velocity range [0, 2] $\rm km\,s^{-1}$, and a possible third fainter filament or separate velocity component in the range [3,4] $\rm km\,s^{-1}$. The PV diagrams of $^{13} \rm CO$ $J=1-0$ line show that the filaments are continuous structures and appear curved towards the clump in the PV diagrams. The curvature in the PV maps can be interpreted as material flowing towards the central clump. Furthermore, the $\rm N_2H^+$ spectra towards pointings 1 to 3 show a clear shift in velocity, which can be interpreted as a streaming motion. Thus, as the clump is found at a crossing of filaments, and since the filaments are connected, at least kinematically to the clump (see Fig. \ref{fig:PV_13}), the clump may have formed through compression by colliding filaments and is still accreting material along the filaments. The curved structures seen in the \textit{Herschel} observations could then have formed by converging flows or by fragmentation within the collapsing cloud. However, at smaller scales, the SMA observations do not show the filamentary structures connecting to the compact sources seen within the clump, but rather the SMA observations seem to trace a shell like structure around the central clump, indicating a more isotropical mass accretion on the compact sources. This combination of large scale filamentary structures and smaller scale isotropic accretion conforms with the hybrid model discussed by \citet{Myers2011}.

A rough estimate can be derived for mass accretion along the substructure S2 onto the central clump, by assuming a simple relation of a core accreting material along a cylindrical structure \citep[following roughly][]{LopezSepulcre2010}

\begin{equation}
\dot{M} = \frac{M_f \hat{V}}{L_f},
\end{equation}
where $M_f$ is the mass of the filamentary structure, $L_f$ is the length of the filament, and $\hat{V}$ is the velocity. For the length and mass of the filament we use values of $M_f = 340 M_\odot$ and $L_f = 1.5$ pc, corresponding to the substructure S2. As for the velocity we use a value $\hat{V} =  1.85$ $\rm km\,s^{-1}$, the difference between the velocity of the $\rm N_2H^+$ hyperfine components from pointings 1 and 3. With the above values, we estimate an accretion rate of $4.3 \times 10^{-4}$ $M_\odot \rm yr^{-1}$. The derived accretion rate is similar to the values reported by \citet{Lu2018} from their observations of high mass star forming regions and similar to the value estimated by \citet{Myers2009}. Thus, the estimated mass accretion rate is sufficient for intermediate-mass and even high-mass star formation.

\subsection{Compact sources}
The central clump is associated with the far-infrared source IRAS 20160+3546. WISE observations show that the central region, which was unresolved by IRAS, contains several near-infrared (NIR) and mid-infrared (MIR) sources within the central region of the clump. The WISE observations at 3.4 $\mu$m and 12 $\mu$m are shown in Fig. \ref{fig:WISE}. Furthermore, the SCUBA-2 maps at 850 $\mu$m show three compact sources within the central clump, but the SMA observations show two continuum sources and three point sources in $^{13} \rm CO$, within the clump. The distances between the three sources are between 0.3 and 0.5 pc, which is close to the Jeans' length \citep{Chandrasekhar_Fermi1953}

\begin{equation}
\lambda_J = \sqrt{\frac{\pi c^2_s}{G \rho}}
\end{equation}
where $c_s$ is the sound speed, $G$ is the gravitational constant and $\rho$ is the density. Assuming a temperature of 15 K, the sound speed is $c_s = 0.21$ $\rm km\,s^{-1}$, and inserting an average value of $\rho$ over the central clump $\rho = 1.65 \times 10^{-20}$ $\rm g \, cm^{-3}$, one gets a Jeans length of $\lambda_J = 0.35$ pc. 

One of the SMA continuum sources, C1, is clearly seen in the $^{13} \rm CO$ and has a clear counterpart in all four WISE channels. The second continuum source, C2, is associated with broader $^{13} \rm CO$ emission and has no clear counterparts in the MIR, although the source is close to an extended source seen in the WISE 3.4 and 4.6 $\mu$m maps. On the other hand, the third source, C3, is not seen in SMA continuum but is seen in SMA $^{13} \rm CO$ observations. The source C3 is not seen in MIR, but at 3.6 $\mu$m and 4.6 $\mu$m there is a point source close to the location of C3. However, this point source is likely a foreground star as it can be identified in the $r, i$, and H$_{\alpha}$ band photometry from the IPHAS survey \citep[The INT Photometric H$_\alpha$ Survey of the Northern Galactic Plane,][]{IPHAS2005}. Furthermore, all three sources have clear counterparts in the SCUBA-2 map at 850 $\mu$m. Thus, it seems that the clump is forming a cluster of at least three stars, all of which are likely in different stages of evolution, judging from the differences in the wavelength coverage. The differences in detections are summarised in Table \ref{table:YSOs}.

\begin{table}
\caption{Summary of detections towards the compact sources C1-C3.}
\centering
\begin{tabular}{c c c c}
\hline\hline
& & &  \\
 & C1 & C2 & C3 \\
\hline
& & & \\

$^{13}$CO 	 		 & Yes & Yes & Yes \\
Continuum 	 		 & Yes & Yes &  No \\
SCUBA-2 	 		 & Yes & Yes & Yes \\
WISE 3.4 $\mu$m  	 & Yes &  No &  No \\
WISE 4.6 $\mu$m		 & Yes &  No &  No \\
WISE 12  $\mu$m		 & Yes &  No &  No \\
WISE 22	 $\mu$m		 & Yes &  No &  No \\
H$_{\alpha}$\tablefootmark{\rm (1)} &  No &  No & Yes \\
$r$\tablefootmark{\rm (1)} 		 	&  No &  No & Yes \\
$i$\tablefootmark{\rm (1)} 		 	&  No &  No & Yes \\

\hline
& & & \\
\end{tabular}
\tablefoot{(1) H$\alpha$, $r$, and $i$ band photometry are from the IPHAS survey.}
\label{table:YSOs}
\end{table}

\subsection{Star formation}

The $\rm H_2O$ maser spectrum detected with the KVN antenna has four velocity components lying within 7 $\rm km\,s^{-1}$ of the average velocity of the dense gas. The central velocity component of the $\rm H_2O$ maser at $\sim -3$ $\rm km\,s^{-1}$ corresponds to the average velocity of the CO lines. Similarly, the $\rm H_2O$ velocity component at $\sim 4$ $\rm km\,s^{-1}$, corresponds to a velocity component seen in the SMA CO spectra. However, the $\rm H_2O$ component at $-7$ $\rm km\,s^{-1}$ is only seen in the SMA CO observations towards the KVN pointing 2. Furthermore, the HCO$^+$ spectrum has an asymmetric line profile which can indicate an outflow, although the asymmetry is weak. On the other hand, the SMA $\rm ^{12}CO$ $J=2-1$ line shows clear inverse P-Cygni profiles towards two of the thee compact sources (see Figs. \ref{fig:SMA_spec_locs} and \ref{fig:SMA12CO}) and the C1 and C3 have broad wings in the line profiles. All these features are signatures of star formation. 

Several studies have reported a correlation between the bolometric luminosity of a source and the isotropic luminosity of a $\rm H_2O$ maser \citep{Bae2011, Brand2003, Furuya2003}. We estimated the isotropic luminosity of the $\rm H_2O$ maser towards pointing 1 to be $L_{22}$ $\sim 2.43 \times 10^{-6}$ $L_\odot$ and derive a bolometric luminosity of 15 $L_\odot$ for the IRAS source. Assuming that the $\rm H_2O$ maser towards pointing 1 and the IRAS source are connected, we compare the above values with Fig. 9 from \citet{Bae2011}, which place our source at the lower end of intermediate mass YSOs, but with a noticeably higher maser luminosity. However, connecting the $\rm H_2O$ maser with one of the compact sources is not straightforward since the beam size of the KVN antenna at 22 GHz is $\sim120\arcsec$, which for pointings 1 and 9 covers the entire central clump. Only one of the compact objects, C3, shows emission at velocities lower than -6 $\rm km\,s^{-1}$ in the SMA $\rm ^{12}CO$ spectra. On the other hand, since the component at $\sim 4$ $\rm km\,s^{-1}$ is only seen towards pointing 1, it is possible that the clump has two maser sources. 

The maser emission, requiring densities of $\sim$10$^9$ cm$^{-3}$ and temperatures around 500 K \citep{Elitzur1989}, is probably caused by shocks in places where protostellar jets hit the surrounding cloud. However, the SMA CO channel maps and spectra show only weak signs of outflows within the clump. On the other hand, because the $\rm H_2O$ profile from pointing 1 has three or four symmetric components, the maser emission can be produced by an accretion disk \citep{Cesaroni1990}. A more detailed study would be required to draw conclusions on the nature of the maser emission, but the low temperature and feeble signs of other stellar feedback inside the central clump indicate that the processes leading to star formation have just recently begun. 

\section{Conclusions}\label{Sect:5}

\renewcommand{\labelitemi}{$-$}

We have studied a star forming region G074.11+00.11. The region comprises a large central clump and several filamentary structures and smaller arms that appear to be interconnected. The central clump is nearly circular, but the densest ridge is clearly elongated from north east to south west and is aligned with a low density filamentary structure. Furthermore, our molecular line observations show that the filamentary structures are kinematically connected with the central clump. Thus, the clump seems to have formed trough a collision of structures at larger scale. The main results of our study are:

\begin{itemize}

\item Based on the \textit{Herschel} and SCUBA-2 observations, the mass of the central clump is $\sim$ 700 $M_\odot$, assuming a distance of 2.3 kpc. The filamentary structures surrounding the central clump have masses in the range [50,340] $M_\odot$ and the total mass of the region is well in excess of 2500 $M_\odot$.

\item The central clump harbours an IRAS point source which in our SCUBA-2 and SMA observations is resolved into three compact objects. Two of these are associated with SMA continuum sources and all three can be identified from SMA CO observations and from the SCUBA-2 observations. However, the three sources behave very differently at WISE wavelengths, a possible indication of different evolutionary stages. We derive a bolometric luminosity of $\sim 15$ $L_\odot$ for the IRAS source and an estimated temperature of $\sim 25$ K.

\item The $^{12} \rm CO$ and $^{13} \rm CO$ lines are broad, with velocities ranging from $-10$ km\,s$^{-1}$ to 5 km\,s$^{-1}$, while the main emission peak is at $-4$ $\rm km\,s^{-1}$. Based on the low-resolution PMO and SMT observations, the CO lines do not show separate velocity components, but rather a smooth velocity gradient over the field. The velocity gradient is also  visible in the position-velocity diagrams drawn through the central clump.

\item The molecular line spectra along the surrounding filamentary structure S1 show a shift in velocity towards the central clump, indicating an ongoing material flow towards the central region of the clump. However, the SMA observations do not show filamentary structures within the clump, thus, any accretion on the compact sources within the clump is likely to proceed isotropically rather than along filaments or through filament fragmentation. The present observations therefore conform with the "hybrid" cluster formation models suggested by \citet{Myers2011} that are isotropic on small scales but filamentary on large scales.

\item We have detected a weak water maser towards two pointings within the central clump, with an estimated isotropic luminosity of $L_{22} = 2.4 \times 10^{-6}$ $L_\odot$. The water maser has four velocity components in the range [$-$10,5] $\rm km\,s^{-1}$, with the central velocity component at $ -4$ $\rm km\,s^{-1}$, which is well aligned with the main CO velocity component. The water maser is likely connected to a weak outflow or generated by an accretion disk. However, because of the large beam size, the exact location of the maser can not be distinguished.

\item The molecular line spectra extracted from the compact sources show inverse P-Cygni profiles characteristic of infall, and asymmetries indicating accretion or low-velocity outflow. Together with water maser emission, these features signify ongoing star formation. However, the present data show no  evidence for high-velocity outflows. We therefore suggest that the region represents a very early stage where feedback from newly born stars has not yet taken effect.

\end{itemize}

In order to properly study the dynamics, and possible accretion along the filaments, a more detailed large scale molecular line study would be beneficial. For example mapping the entire SCUBA-2 region with high density tracers, such as $\rm N_2H^+$, could be used to study and model the region in much higher detail.

\begin{acknowledgements}
MS, JH, and MJ acknowledge  the support of the Academy of Finland Grant No. 285769. This work has made use of data from the European Space Agency (ESA) mission {\it Gaia} (\url{https://www.cosmos.esa.int/gaia}), processed by the {\it Gaia} Data Processing and Analysis Consortium (DPAC, \url{https://www.cosmos.esa.int/web/gaia/dpac/consortium}). Funding for the DPAC
has been provided by national institutions, in particular the institutions participating in the {\it Gaia} Multilateral Agreement. The data presented in this paper are based on the ESO-ARO programme ID 196.C-0999(A). Ke Wang acknowledges support by the National Key Research and Development Program of China (2017YFA0402702), the National Science Foundation of China (11721303),
and the starting grant at the Kavli Institute for Astronomy and Astrophysics, Peking University (7101502016). 

\end{acknowledgements}

\bibliographystyle{aa}
\bibliography{bibli}

\begin{thebibliography}{68}
\expandafter\ifx\csname natexlab\endcsname\relax\def\natexlab#1{#1}\fi

\bibitem[{{Alonso-Albi} {et~al.}(2009){Alonso-Albi}, {Fuente}, {Bachiller},
  {Neri}, {Planesas}, {Testi}, {Bern{\'e}}, \& {Joblin}}]{Alonso-Albi2009}
{Alonso-Albi}, T., {Fuente}, A., {Bachiller}, R., {et~al.} 2009, \aap, 497, 117

\bibitem[{{Andrae} {et~al.}(2018){Andrae}, {Fouesneau}, {Creevey}, {Ordenovic},
  {Mary}, {Burlacu}, {Chaoul}, {Jean-Antoine-Piccolo}, {Kordopatis}, {Korn},
  {Lebreton}, {Panem}, {Pichon}, {Th{\'e}venin}, {Walmsley}, \&
  {Bailer-Jones}}]{Andrae2018}
{Andrae}, R., {Fouesneau}, M., {Creevey}, O., {et~al.} 2018, \aap, 616, A8

\bibitem[{{Andr{\'e}} {et~al.}(2014){Andr{\'e}}, {Di Francesco},
  {Ward-Thompson}, {Inutsuka}, {Pudritz}, \& {Pineda}}]{Andre_PPVI2014}
{Andr{\'e}}, P., {Di Francesco}, J., {Ward-Thompson}, D., {et~al.} 2014,
  Protostars and Planets VI, 27

\bibitem[{{Bae} {et~al.}(2011){Bae}, {Kim}, {Youn}, {Kim}, {Byun}, {Kang}, \&
  {Oh}}]{Bae2011}
{Bae}, J.-H., {Kim}, K.-T., {Youn}, S.-Y., {et~al.} 2011, \apjs, 196, 21

\bibitem[{{Bailer-Jones}(2015)}]{Bailer-Jones2015}
{Bailer-Jones}, C.~A.~L. 2015, \pasp, 127, 994

\bibitem[{{Beckwith} {et~al.}(1990){Beckwith}, {Sargent}, {Chini}, \&
  {Guesten}}]{Beckwith1990}
{Beckwith}, S.~V.~W., {Sargent}, A.~I., {Chini}, R.~S., \& {Guesten}, R. 1990,
  \aj, 99, 924

\bibitem[{{Brand} {et~al.}(2003){Brand}, {Cesaroni}, {Comoretto}, {Felli},
  {Palagi}, {Palla}, \& {Valdettaro}}]{Brand2003}
{Brand}, J., {Cesaroni}, R., {Comoretto}, G., {et~al.} 2003, \aap, 407, 573

\bibitem[{{Busquet} {et~al.}(2013){Busquet}, {Zhang}, {Palau}, {Liu},
  {S{\'a}nchez-Monge}, {Estalella}, {Ho}, {de Gregorio-Monsalvo}, {Pillai},
  {Wyrowski}, {Girart}, {Santos}, \& {Franco}}]{Busquet2013}
{Busquet}, G., {Zhang}, Q., {Palau}, A., {et~al.} 2013, \apjl, 764, L26

\bibitem[{{Cesaroni}(1990)}]{Cesaroni1990}
{Cesaroni}, R. 1990, \aap, 233, 513

\bibitem[{{Chandler} \& {Richer}(2000)}]{Chandler2000}
{Chandler}, C.~J. \& {Richer}, J.~S. 2000, \apj, 530, 851

\bibitem[{{Chandrasekhar} \& {Fermi}(1953)}]{Chandrasekhar_Fermi1953}
{Chandrasekhar}, S. \& {Fermi}, E. 1953, \apj, 118, 116

\bibitem[{{Chapin} {et~al.}(2013){Chapin}, {Berry}, {Gibb}, {Jenness}, {Scott},
  {Tilanus}, {Economou}, \& {Holland}}]{Chapin2013}
{Chapin}, E.~L., {Berry}, D.~S., {Gibb}, A.~G., {et~al.} 2013, \mnras, 430,
  2545

\bibitem[{{Drew} {et~al.}(2005){Drew}, {Greimel}, {Irwin}, {Aungwerojwit},
  {Barlow}, {Corradi}, {Drake}, {G{\"a}nsicke}, {Groot}, {Hales}, {Hopewell},
  {Irwin}, {Knigge}, {Leisy}, {Lennon}, {Mampaso}, {Masheder}, {Matsuura},
  {Morales-Rueda}, {Morris}, {Parker}, {Phillipps}, {Rodriguez-Gil}, {Roelofs},
  {Skillen}, {Sokoloski}, {Steeghs}, {Unruh}, {Viironen}, {Vink}, {Walton},
  {Witham}, {Wright}, {Zijlstra}, \& {Zurita}}]{IPHAS2005}
{Drew}, J.~E., {Greimel}, R., {Irwin}, M.~J., {et~al.} 2005, \mnras, 362, 753

\bibitem[{{Eden} {et~al.}(2019){Eden}, {Liu}, {Kim}, {Juvela}, {Liu},
  {Tatematsu}, {Francesco}, {Wang}, {Wu}, {Thompson}, {Fuller}, {Li},
  {Ristorcelli}, {Kang}, {Hirano}, {Johnstone}, {Lin}, {He}, {Koch},
  {Sanhueza}, {Qin}, {Zhang}, {Goldsmith}, {Evans}, {Yuan}, {Zhang}, {White},
  {Choi}, {Lee}, {Toth}, {Mairs}, {Yi}, {Tang}, {Soam}, {Peretto}, {Samal},
  {Fich}, {Parsons}, {Malinen}, {Bendo}, {Rivera-Ingraham}, {Liu},
  {Wouterloot}, {Li}, {Qian}, {Rawlings}, {Rawlings}, {Feng}, {Wang}, {Li},
  {Liu}, {Luo}, {Marston}, {Pattle}, {Pelkonen}, {Rigby}, {Zahorecz}, {Zhang},
  {B{\H{o}}gner}, {Aikawa}, {Akhter}, {Alina}, {Bell}, {Bernard}, {Blain},
  {Bronfman}, {Byun}, {Chapman}, {Chen}, {Chen}, {Chen}, {Chen}, {Chen},
  {Chrysostomou}, {Chu}, {Chung}, {Cornu}, {Cosentino}, {Cunningham}, {Demyk},
  {Drabek-Maunder}, {Doi}, {Eswaraiah}, {Falgarone}, {Feh{\'e}r}, {Fraser},
  {Friberg}, {Garay}, {Ge}, {Gear}, {Greaves}, {Guan}, {Harvey-Smith},
  {Hasegawa}, {He}, {Henkel}, {Hirota}, {Holland}, {Hughes}, {Jarken}, {Ji},
  {Jimenez-Serra}, {Kang}, {Kawabata}, {Kim}, {Kim}, {Kim}, {Kim}, {Koo},
  {Kwon}, {Kuan}, {Lacaille}, {Lai}, {Lee}, {Lee}, {Lee}, {Li}, {Lo}, {Lopez},
  {Lu}, {Lyo}, {Mardones}, {McGehee}, {Meng}, {Montier}, {Montillaud}, {Moore},
  {Morata}, {Moriarty-Schieven}, {Ohashi}, {Pak}, {Park}, {Paladini}, {Pech},
  {Qiu}, {Ren}, {Richer}, {Sakai}, {Shang}, {Shinnaga}, {Stamatellos}, {Tang},
  {Traficante}, {Vastel}, {Viti}, {Walsh}, {Wang}, {Wang}, {Ward-Thompson},
  {Whitworth}, {Wilson}, {Xu}, {Yang}, {Yuan}, {Yuan}, {Zavagno}, {Zhang},
  {Zhang}, {Zhang}, {Zhou}, {Zhou}, {Zhu}, \& {Zuo}}]{Eden2019}
{Eden}, D.~J., {Liu}, T., {Kim}, K.-T., {et~al.} 2019, \mnras, 485, 2895

\bibitem[{{Elitzur} {et~al.}(1989){Elitzur}, {Hollenbach}, \&
  {McKee}}]{Elitzur1989}
{Elitzur}, M., {Hollenbach}, D.~J., \& {McKee}, C.~F. 1989, \apj, 346, 983

\bibitem[{{Fuller} {et~al.}(2005){Fuller}, {Williams}, \&
  {Sridharan}}]{Fuller2005}
{Fuller}, G.~A., {Williams}, S.~J., \& {Sridharan}, T.~K. 2005, \aap, 442, 949

\bibitem[{{Furuya} {et~al.}(2003){Furuya}, {Kitamura}, {Wootten}, {Claussen},
  \& {Kawabe}}]{Furuya2003}
{Furuya}, R.~S., {Kitamura}, Y., {Wootten}, A., {Claussen}, M.~J., \& {Kawabe},
  R. 2003, \apjs, 144, 71

\bibitem[{{Gaia Collaboration} {et~al.}(2018){Gaia Collaboration}, {Brown},
  {Vallenari}, {Prusti}, {de Bruijne}, {Babusiaux}, {Bailer-Jones}, {Biermann},
  {Evans}, {Eyer}, \& et~al.}]{GAIA_DR2_2018}
{Gaia Collaboration}, {Brown}, A.~G.~A., {Vallenari}, A., {et~al.} 2018, \aap,
  616, A1

\bibitem[{{Gaia Collaboration} {et~al.}(2016){Gaia Collaboration}, {Prusti},
  {de Bruijne}, {Brown}, {Vallenari}, {Babusiaux}, {Bailer-Jones}, {Bastian},
  {Biermann}, {Evans}, {Eyer}, {Jansen}, {Jordi}, {Klioner}, {Lammers},
  {Lindegren}, {Luri}, {Mignard}, {Milligan}, {Panem}, {Poinsignon},
  {Pourbaix}, {Randich}, {Sarri}, {Sartoretti}, {Siddiqui}, {Soubiran},
  {Valette}, {van Leeuwen}, {Walton}, {Aerts}, {Arenou}, {Cropper}, {Drimmel},
  {H{\o}g}, {Katz}, {Lattanzi}, {O'Mullane}, {Grebel}, {Holland}, {Huc},
  {Passot}, {Bramante}, {Cacciari}, {Casta{\~n}eda}, {Chaoul}, {Cheek}, {De
  Angeli}, {Fabricius}, {Guerra}, {Hern{\'a}ndez}, {Jean-Antoine-Piccolo},
  {Masana}, {Messineo}, {Mowlavi}, {Nienartowicz}, {Ord{\'o}{\~n}ez-Blanco},
  {Panuzzo}, {Portell}, {Richards}, {Riello}, {Seabroke}, {Tanga},
  {Th{\'e}venin}, {Torra}, {Els}, {Gracia-Abril}, {Comoretto},
  {Garcia-Reinaldos}, {Lock}, {Mercier}, {Altmann}, {Andrae}, {Astraatmadja},
  {Bellas-Velidis}, {Benson}, {Berthier}, {Blomme}, {Busso}, {Carry},
  {Cellino}, {Clementini}, {Cowell}, {Creevey}, {Cuypers}, {Davidson}, {De
  Ridder}, {de Torres}, {Delchambre}, {Dell'Oro}, {Ducourant}, {Fr{\'e}mat},
  {Garc{\'\i}a-Torres}, {Gosset}, {Halbwachs}, {Hambly}, {Harrison}, {Hauser},
  {Hestroffer}, {Hodgkin}, {Huckle}, {Hutton}, {Jasniewicz}, {Jordan},
  {Kontizas}, {Korn}, {Lanzafame}, {Manteiga}, {Moitinho}, {Muinonen},
  {Osinde}, {Pancino}, {Pauwels}, {Petit}, {Recio-Blanco}, {Robin}, {Sarro},
  {Siopis}, {Smith}, {Smith}, {Sozzetti}, {Thuillot}, {van Reeven}, {Viala},
  {Abbas}, {Abreu Aramburu}, {Accart}, {Aguado}, {Allan}, {Allasia},
  {Altavilla}, {{\'A}lvarez}, {Alves}, {Anderson}, {Andrei}, {Anglada Varela},
  {Antiche}, {Antoja}, {Ant{\'o}n}, {Arcay}, {Atzei}, {Ayache}, {Bach},
  {Baker}, {Balaguer-N{\'u}{\~n}ez}, {Barache}, {Barata}, {Barbier}, {Barblan},
  {Baroni}, {Barrado y Navascu{\'e}s}, {Barros}, {Barstow}, {Becciani},
  {Bellazzini}, {Bellei}, {Bello Garc{\'\i}a}, {Belokurov}, {Bendjoya},
  {Berihuete}, {Bianchi}, {Bienaym{\'e}}, {Billebaud}, {Blagorodnova},
  {Blanco-Cuaresma}, {Boch}, {Bombrun}, {Borrachero}, {Bouquillon}, {Bourda},
  {Bouy}, {Bragaglia}, {Breddels}, {Brouillet}, {Br{\"u}semeister},
  {Bucciarelli}, {Budnik}, {Burgess}, {Burgon}, {Burlacu}, {Busonero}, {Buzzi},
  {Caffau}, {Cambras}, {Campbell}, {Cancelliere}, {Cantat-Gaudin}, {Carlucci},
  {Carrasco}, {Castellani}, {Charlot}, {Charnas}, {Charvet}, {Chassat},
  {Chiavassa}, {Clotet}, {Cocozza}, {Collins}, {Collins}, {Costigan}, {Crifo},
  {Cross}, {Crosta}, {Crowley}, {Dafonte}, {Damerdji}, {Dapergolas}, {David},
  {David}, {De Cat}, {de Felice}, {de Laverny}, {De Luise}, {De March}, {de
  Martino}, {de Souza}, {Debosscher}, {del Pozo}, {Delbo}, {Delgado},
  {Delgado}, {di Marco}, {Di Matteo}, {Diakite}, {Distefano}, {Dolding}, {Dos
  Anjos}, {Drazinos}, {Dur{\'a}n}, {Dzigan}, {Ecale}, {Edvardsson}, {Enke},
  {Erdmann}, {Escolar}, {Espina}, {Evans}, {Eynard Bontemps}, {Fabre},
  {Fabrizio}, {Faigler}, {Falc{\~a}o}, {Farr{\`a}s Casas}, {Faye}, {Federici},
  {Fedorets}, {Fern{\'a}ndez-Hern{\'a}ndez}, {Fernique}, {Fienga}, {Figueras},
  {Filippi}, {Findeisen}, {Fonti}, {Fouesneau}, {Fraile}, {Fraser}, {Fuchs},
  {Furnell}, {Gai}, {Galleti}, {Galluccio}, {Garabato}, {Garc{\'\i}a-Sedano},
  {Gar{\'e}}, {Garofalo}, {Garralda}, {Gavras}, {Gerssen}, {Geyer}, {Gilmore},
  {Girona}, {Giuffrida}, {Gomes}, {Gonz{\'a}lez-Marcos},
  {Gonz{\'a}lez-N{\'u}{\~n}ez}, {Gonz{\'a}lez-Vidal}, {Granvik}, {Guerrier},
  {Guillout}, {Guiraud}, {G{\'u}rpide}, {Guti{\'e}rrez-S{\'a}nchez}, {Guy},
  {Haigron}, {Hatzidimitriou}, {Haywood}, {Heiter}, {Helmi}, {Hobbs},
  {Hofmann}, {Holl}, {Holland }, {Hunt}, {Hypki}, {Icardi}, {Irwin}, {Jevardat
  de Fombelle}, {Jofr{\'e}}, {Jonker}, {Jorissen}, {Julbe}, {Karampelas},
  {Kochoska}, {Kohley}, {Kolenberg}, {Kontizas}, {Koposov}, {Kordopatis},
  {Koubsky}, {Kowalczyk}, {Krone-Martins}, {Kudryashova}, {Kull}, {Bachchan},
  {Lacoste-Seris}, {Lanza}, {Lavigne}, {Le Poncin-Lafitte}, {Lebreton},
  {Lebzelter}, {Leccia}, {Leclerc}, {Lecoeur-Taibi}, {Lemaitre}, {Lenhardt},
  {Leroux}, {Liao}, {Licata}, {Lindstr{\o}m}, {Lister}, {Livanou}, {Lobel},
  {L{\"o}ffler}, {L{\'o}pez}, {Lopez-Lozano}, {Lorenz}, {Loureiro},
  {MacDonald}, {Magalh{\~a}es Fernandes}, {Managau}, {Mann}, {Mantelet},
  {Marchal}, {Marchant}, {Marconi}, {Marie}, {Marinoni}, {Marrese},
  {Marschalk{\'o}}, {Marshall}, {Mart{\'\i}n-Fleitas}, {Martino}, {Mary},
  {Matijevi{\v{c}}}, {Mazeh}, {McMillan}, {Messina}, {Mestre}, {Michalik},
  {Millar}, {Miranda}, {Molina}, {Molinaro}, {Molinaro}, {Moln{\'a}r},
  {Moniez}, {Montegriffo}, {Monteiro}, {Mor}, {Mora}, {Morbidelli}, {Morel},
  {Morgenthaler}, {Morley}, {Morris}, {Mulone}, {Muraveva}, {Musella},
  {Narbonne}, {Nelemans}, {Nicastro}, {Noval}, {Ord{\'e}novic},
  {Ordieres-Mer{\'e}}, {Osborne}, {Pagani}, {Pagano}, {Pailler}, {Palacin},
  {Palaversa}, {Parsons}, {Paulsen}, {Pecoraro}, {Pedrosa}, {Pentik{\"a}inen},
  {Pereira}, {Pichon}, {Piersimoni}, {Pineau}, {Plachy}, {Plum}, {Poujoulet},
  {Pr{\v{s}}a}, {Pulone}, {Ragaini}, {Rago}, {Rambaux}, {Ramos-Lerate},
  {Ranalli}, {Rauw}, {Read}, {Regibo}, {Renk}, {Reyl{\'e}}, {Ribeiro},
  {Rimoldini}, {Ripepi}, {Riva}, {Rixon}, {Roelens}, {Romero-G{\'o}mez},
  {Rowell}, {Royer}, {Rudolph}, {Ruiz-Dern}, {Sadowski}, {Sagrist{\`a}
  Sell{\'e}s}, {Sahlmann}, {Salgado}, {Salguero}, {Sarasso}, {Savietto},
  {Schnorhk}, {Schultheis}, {Sciacca}, {Segol}, {Segovia}, {Segransan},
  {Serpell}, {Shih}, {Smareglia}, {Smart}, {Smith}, {Solano}, {Solitro},
  {Sordo}, {Soria Nieto}, {Souchay}, {Spagna}, {Spoto}, {Stampa}, {Steele},
  {Steidelm{\"u}ller}, {Stephenson}, {Stoev}, {Suess}, {S{\"u}veges}, {Surdej},
  {Szabados}, {Szegedi-Elek}, {Tapiador}, {Taris}, {Tauran}, {Taylor},
  {Teixeira}, {Terrett}, {Tingley}, {Trager}, {Turon}, {Ulla}, {Utrilla},
  {Valentini}, {van Elteren}, {Van Hemelryck}, {van Leeuwen}, {Varadi},
  {Vecchiato}, {Veljanoski}, {Via}, {Vicente}, {Vogt}, {Voss}, {Votruba},
  {Voutsinas}, {Walmsley}, {Weiler}, {Weingrill}, {Werner}, {Wevers},
  {Whitehead}, {Wyrzykowski}, {Yoldas}, {{\v{Z}}erjal}, {Zucker}, {Zurbach},
  {Zwitter}, {Alecu}, {Allen}, {Allende Prieto}, {Amorim},
  {Anglada-Escud{\'e}}, {Arsenijevic}, {Azaz}, {Balm}, {Beck}, {Bernstein},
  {Bigot}, {Bijaoui}, {Blasco}, {Bonfigli}, {Bono}, {Boudreault}, {Bressan},
  {Brown}, {Brunet}, {Bunclark}, {Buonanno}, {Butkevich}, {Carret}, {Carrion},
  {Chemin}, {Ch{\'e}reau}, {Corcione}, {Darmigny}, {de Boer}, {de Teodoro}, {de
  Zeeuw}, {Delle Luche}, {Domingues}, {Dubath}, {Fodor}, {Fr{\'e}zouls},
  {Fries}, {Fustes}, {Fyfe}, {Gallardo}, {Gallegos}, {Gardiol}, {Gebran},
  {Gomboc}, {G{\'o}mez}, {Grux}, {Gueguen}, {Heyrovsky}, {Hoar}, {Iannicola},
  {Isasi Parache}, {Janotto}, {Joliet}, {Jonckheere}, {Keil}, {Kim},
  {Klagyivik}, {Klar}, {Knude}, {Kochukhov}, {Kolka}, {Kos}, {Kutka}, {Lainey},
  {LeBouquin}, {Liu}, {Loreggia}, {Makarov}, {Marseille}, {Martayan},
  {Martinez-Rubi}, {Massart}, {Meynadier}, {Mignot}, {Munari}, {Nguyen},
  {Nordlander}, {Ocvirk}, {O'Flaherty}, {Olias Sanz}, {Ortiz}, {Osorio},
  {Oszkiewicz}, {Ouzounis}, {Palmer}, {Park}, {Pasquato}, {Peltzer}, {Peralta},
  {P{\'e}turaud}, {Pieniluoma}, {Pigozzi}, {Poels}, {Prat}, {Prod'homme},
  {Raison}, {Rebordao}, {Risquez}, {Rocca-Volmerange}, {Rosen}, {Ruiz-Fuertes},
  {Russo}, {Sembay}, {Serraller Vizcaino}, {Short}, {Siebert}, {Silva},
  {Sinachopoulos}, {Slezak}, {Soffel}, {Sosnowska}, {Strai{\v{z}}ys}, {ter
  Linden}, {Terrell}, {Theil}, {Tiede}, {Troisi}, {Tsalmantza}, {Tur},
  {Vaccari}, {Vachier}, {Valles}, {Van Hamme}, {Veltz}, {Virtanen}, {Wallut},
  {Wichmann}, {Wilkinson}, {Ziaeepour}, \& {Zschocke}}]{GAIA2016}
{Gaia Collaboration}, {Prusti}, T., {de Bruijne}, J.~H.~J., {et~al.} 2016,
  \aap, 595, A1

\bibitem[{{Galv{\'a}n-Madrid} {et~al.}(2010){Galv{\'a}n-Madrid}, {Zhang},
  {Keto}, {Ho}, {Zapata}, {Rodr{\'{\i}}guez}, {Pineda}, \&
  {V{\'a}zquez-Semadeni}}]{Galvan-Madrid2010}
{Galv{\'a}n-Madrid}, R., {Zhang}, Q., {Keto}, E., {et~al.} 2010, \apj, 725, 17

\bibitem[{{Giannetti} {et~al.}(2013){Giannetti}, {Brand}, {S{\'a}nchez-Monge},
  {Fontani}, {Cesaroni}, {Beltr{\'a}n}, {Molinari}, {Dodson}, \&
  {Rioja}}]{Giannetti2013}
{Giannetti}, A., {Brand}, J., {S{\'a}nchez-Monge}, {\'A}., {et~al.} 2013, \aap,
  556, A16

\bibitem[{{Griffin} {et~al.}(2010){Griffin}, {Abergel}, {Abreu}, {Ade},
  {Andr{\'e}}, {Augueres}, {Babbedge}, {Bae}, {Baillie}, {Baluteau}, {Barlow},
  {Bendo}, {Benielli}, {Bock}, {Bonhomme}, {Brisbin}, {Brockley-Blatt},
  {Caldwell}, {Cara}, {Castro-Rodriguez}, {Cerulli}, {Chanial}, {Chen},
  {Clark}, {Clements}, {Clerc}, {Coker}, {Communal}, {Conversi}, {Cox},
  {Crumb}, {Cunningham}, {Daly}, {Davis}, {de Antoni}, {Delderfield}, {Devin},
  {di Giorgio}, {Didschuns}, {Dohlen}, {Donati}, {Dowell}, {Dowell}, {Duband},
  {Dumaye}, {Emery}, {Ferlet}, {Ferrand}, {Fontignie}, {Fox}, {Franceschini},
  {Frerking}, {Fulton}, {Garcia}, {Gastaud}, {Gear}, {Glenn}, {Goizel},
  {Griffin}, {Grundy}, {Guest}, {Guillemet}, {Hargrave}, {Harwit}, {Hastings},
  {Hatziminaoglou}, {Herman}, {Hinde}, {Hristov}, {Huang}, {Imhof}, {Isaak},
  {Israelsson}, {Ivison}, {Jennings}, {Kiernan}, {King}, {Lange}, {Latter},
  {Laurent}, {Laurent}, {Leeks}, {Lellouch}, {Levenson}, {Li}, {Li},
  {Lilienthal}, {Lim}, {Liu}, {Lu}, {Madden}, {Mainetti}, {Marliani}, {McKay},
  {Mercier}, {Molinari}, {Morris}, {Moseley}, {Mulder}, {Mur}, {Naylor},
  {Nguyen}, {O'Halloran}, {Oliver}, {Olofsson}, {Olofsson}, {Orfei}, {Page},
  {Pain}, {Panuzzo}, {Papageorgiou}, {Parks}, {Parr-Burman}, {Pearce},
  {Pearson}, {P{\'e}rez-Fournon}, {Pinsard}, {Pisano}, {Podosek}, {Pohlen},
  {Polehampton}, {Pouliquen}, {Rigopoulou}, {Rizzo}, {Roseboom}, {Roussel},
  {Rowan-Robinson}, {Rownd}, {Saraceno}, {Sauvage}, {Savage}, {Savini},
  {Sawyer}, {Scharmberg}, {Schmitt}, {Schneider}, {Schulz}, {Schwartz},
  {Shafer}, {Shupe}, {Sibthorpe}, {Sidher}, {Smith}, {Smith}, {Smith},
  {Spencer}, {Stobie}, {Sudiwala}, {Sukhatme}, {Surace}, {Stevens}, {Swinyard},
  {Trichas}, {Tourette}, {Triou}, {Tseng}, {Tucker}, {Turner}, {Vaccari},
  {Valtchanov}, {Vigroux}, {Virique}, {Voellmer}, {Walker}, {Ward}, {Waskett},
  {Weilert}, {Wesson}, {White}, {Whitehouse}, {Wilson}, {Winter}, {Woodcraft},
  {Wright}, {Xu}, {Zavagno}, {Zemcov}, {Zhang}, \& {Zonca}}]{Griffin2010}
{Griffin}, M.~J., {Abergel}, A., {Abreu}, A., {et~al.} 2010, \aap, 518, L3

\bibitem[{{Hildebrand}(1983)}]{Hildebrand1983}
{Hildebrand}, R.~H. 1983, \qjras, 24, 267

\bibitem[{{Ho} {et~al.}(2004){Ho}, {Moran}, \& {Lo}}]{Ho_sma}
{Ho}, P.~T.~P., {Moran}, J.~M., \& {Lo}, K.~Y. 2004, \apjl, 616, L1

\bibitem[{{Holland} {et~al.}(2013){Holland}, {Bintley}, {Chapin},
  {Chrysostomou}, {Davis}, {Dempsey}, {Duncan}, {Fich}, {Friberg}, {Halpern},
  {Irwin}, {Jenness}, {Kelly}, {MacIntosh}, {Robson}, {Scott}, {Ade},
  {Atad-Ettedgui}, {Berry}, {Craig}, {Gao}, {Gibb}, {Hilton}, {Hollister},
  {Kycia}, {Lunney}, {McGregor}, {Montgomery}, {Parkes}, {Tilanus}, {Ullom},
  {Walther}, {Walton}, {Woodcraft}, {Amiri}, {Atkinson}, {Burger}, {Chuter},
  {Coulson}, {Doriese}, {Dunare}, {Economou}, {Niemack}, {Parsons},
  {Reintsema}, {Sibthorpe}, {Smail}, {Sudiwala}, \& {Thomas}}]{Holland2013}
{Holland}, W.~S., {Bintley}, D., {Chapin}, E.~L., {et~al.} 2013, \mnras, 430,
  2513

\bibitem[{{Juvela} {et~al.}(2015){Juvela}, {Ristorcelli}, {Marshall},
  {Montillaud}, {Pelkonen}, {Ysard}, {McGehee}, {Paladini}, {Pagani},
  {Malinen}, {Rivera-Ingraham}, {Lef{\`e}vre}, {T{\'o}th}, {Montier},
  {Bernard}, \& {Martin}}]{Juvela2015V}
{Juvela}, M., {Ristorcelli}, I., {Marshall}, D.~J., {et~al.} 2015, \aap, 584,
  A93

\bibitem[{{Karr} \& {Martin}(2003)}]{Karr2003}
{Karr}, J.~L. \& {Martin}, P.~G. 2003, \apj, 595, 900

\bibitem[{{Kim} {et~al.}(2018){Kim}, {Kim}, \& {Park}}]{Kim2018}
{Kim}, C.-H., {Kim}, K.-T., \& {Park}, Y.-S. 2018, \apjs, 236, 31

\bibitem[{{Lee} {et~al.}(2000){Lee}, {Mundy}, {Reipurth}, {Ostriker}, \&
  {Stone}}]{Lee2000}
{Lee}, C.-F., {Mundy}, L.~G., {Reipurth}, B., {Ostriker}, E.~C., \& {Stone},
  J.~M. 2000, \apj, 542, 925

\bibitem[{{Lee} \& {Chen}(2007)}]{Lee2007}
{Lee}, H.-T. \& {Chen}, W.~P. 2007, \apj, 657, 884

\bibitem[{{Liu} {et~al.}(2015){Liu}, {Galv{\'a}n-Madrid}, {Jim{\'e}nez-Serra},
  {Rom{\'a}n-Z{\'u}{\~n}iga}, {Zhang}, {Li}, \& {Chen}}]{Liu_H2015}
{Liu}, H.~B., {Galv{\'a}n-Madrid}, R., {Jim{\'e}nez-Serra}, I., {et~al.} 2015,
  \apj, 804, 37

\bibitem[{{Liu} {et~al.}(2012{\natexlab{a}}){Liu}, {Jim{\'e}nez-Serra}, {Ho},
  {Chen}, {Zhang}, \& {Li}}]{Liu_H2012}
{Liu}, H.~B., {Jim{\'e}nez-Serra}, I., {Ho}, P.~T.~P., {et~al.}
  2012{\natexlab{a}}, \apj, 756, 10

\bibitem[{{Liu} {et~al.}(2012{\natexlab{b}}){Liu}, {Quintana-Lacaci}, {Wang},
  {Ho}, {Li}, {Zhang}, \& {Zhang}}]{LiuH2012}
{Liu}, H.~B., {Quintana-Lacaci}, G., {Wang}, K., {et~al.} 2012{\natexlab{b}},
  \apj, 745, 61

\bibitem[{{Liu} {et~al.}(2018){Liu}, {Kim}, {Juvela}, {Wang}, {Tatematsu}, {Di
  Francesco}, {Liu}, {Wu}, {Thompson}, {Fuller}, {Eden}, {Li}, {Ristorcelli},
  {Kang}, {Lin}, {Johnstone}, {He}, {Koch}, {Sanhueza}, {Qin}, {Zhang},
  {Hirano}, {Goldsmith}, {Evans}, {White}, {Choi}, {Lee}, {Toth}, {Mairs},
  {Yi}, {Tang}, {Soam}, {Peretto}, {Samal}, {Fich}, {Parsons}, {Yuan}, {Zhang},
  {Malinen}, {Bendo}, {Rivera-Ingraham}, {Liu}, {Wouterloot}, {Li}, {Qian},
  {Rawlings}, {Rawlings}, {Feng}, {Aikawa}, {Akhter}, {Alina}, {Bell},
  {Bernard}, {Blain}, {B{\H o}gner}, {Bronfman}, {Byun}, {Chapman}, {Chen},
  {Chen}, {Chen}, {Chen}, {Chen}, {Chrysostomou}, {Cosentino}, {Cunningham},
  {Demyk}, {Drabek-Maunder}, {Doi}, {Eswaraiah}, {Falgarone}, {Feh{\'e}r},
  {Fraser}, {Friberg}, {Garay}, {Ge}, {Gear}, {Greaves}, {Guan},
  {Harvey-Smith}, {HASEGAWA}, {Hatchell}, {He}, {Henkel}, {Hirota}, {Holland},
  {Hughes}, {Jarken}, {Ji}, {Jimenez-Serra}, {Kang}, {Kawabata}, {Kim}, {Kim},
  {Kim}, {Kim}, {Koo}, {Kwon}, {Kuan}, {Lacaille}, {Lai}, {Lee}, {Lee}, {Lee},
  {Li}, {Li}, {Lo}, {Lopez}, {Lu}, {Lyo}, {Mardones}, {Marston}, {McGehee},
  {Meng}, {Montier}, {Montillaud}, {Moore}, {Morata}, {Moriarty-Schieven},
  {Ohashi}, {Pak}, {Park}, {Paladini}, {Pattle}, {Pech}, {Pelkonen}, {Qiu},
  {Ren}, {Richer}, {Saito}, {Sakai}, {Shang}, {Shinnaga}, {Stamatellos},
  {Tang}, {Traficante}, {Vastel}, {Viti}, {Walsh}, {Wang}, {Wang}, {Wang},
  {Ward-Thompson}, {Whitworth}, {Xu}, {Yang}, {Yang}, {Yuan}, {Zavagno},
  {Zhang}, {Zhang}, {Zhou}, {Zhou}, {Zhu}, {Zuo}, \& {Zhang}}]{Liu2018TOPSCOPE}
{Liu}, T., {Kim}, K.-T., {Juvela}, M., {et~al.} 2018, \apjs, 234, 28

\bibitem[{{Liu} {et~al.}(2012{\natexlab{c}}){Liu}, {Wu}, \& {Zhang}}]{Liu2012}
{Liu}, T., {Wu}, Y., \& {Zhang}, H. 2012{\natexlab{c}}, \apjs, 202, 4

\bibitem[{{Liu} {et~al.}(2013){Liu}, {Wu}, \& {Zhang}}]{Liu2013}
{Liu}, T., {Wu}, Y., \& {Zhang}, H. 2013, \apjl, 775, L2

\bibitem[{{Liu} {et~al.}(2016){Liu}, {Zhang}, {Kim}, {Wu}, {Lee}, {Goldsmith},
  {Li}, {Liu}, {Chen}, {Tatematsu}, {Wang}, {Lee}, {Qin}, {Mardones}, \&
  {Cho}}]{Liu2016}
{Liu}, T., {Zhang}, Q., {Kim}, K.-T., {et~al.} 2016, \apj, 824, 31

\bibitem[{{L{\'o}pez-Sepulcre} {et~al.}(2010){L{\'o}pez-Sepulcre}, {Cesaroni},
  \& {Walmsley}}]{LopezSepulcre2010}
{L{\'o}pez-Sepulcre}, A., {Cesaroni}, R., \& {Walmsley}, C.~M. 2010, \aap, 517,
  A66

\bibitem[{{Lu} {et~al.}(2018){Lu}, {Zhang}, {Liu}, {Sanhueza}, {Tatematsu},
  {Feng}, {Smith}, {Myers}, {Sridharan}, \& {Gu}}]{Lu2018}
{Lu}, X., {Zhang}, Q., {Liu}, H.~B., {et~al.} 2018, \apj, 855, 9

\bibitem[{{Luri} {et~al.}(2018){Luri}, {Brown}, {Sarro}, {Arenou},
  {Bailer-Jones}, {Castro-Ginard}, {de Bruijne}, {Prusti}, {Babusiaux}, \&
  {Delgado}}]{Luri2018}
{Luri}, X., {Brown}, A.~G.~A., {Sarro}, L.~M., {et~al.} 2018, \aap, 616, A9

\bibitem[{{MacLaren} {et~al.}(1988){MacLaren}, {Richardson}, \&
  {Wolfendale}}]{MacLaren}
{MacLaren}, I., {Richardson}, K.~M., \& {Wolfendale}, A.~W. 1988, \apj, 333,
  821

\bibitem[{{Mardones} {et~al.}(1997){Mardones}, {Myers}, {Tafalla}, {Wilner},
  {Bachiller}, \& {Garay}}]{Mardones1997}
{Mardones}, D., {Myers}, P.~C., {Tafalla}, M., {et~al.} 1997, \apj, 489, 719

\bibitem[{{Marshall} {et~al.}(2006){Marshall}, {Robin}, {Reyl{\'e}},
  {Schultheis}, \& {Picaud}}]{Marshall2006}
{Marshall}, D.~J., {Robin}, A.~C., {Reyl{\'e}}, C., {Schultheis}, M., \&
  {Picaud}, S. 2006, \aap, 453, 635

\bibitem[{{Meng} {et~al.}(2013){Meng}, {Wu}, \& {Liu}}]{Meng2013}
{Meng}, F., {Wu}, Y., \& {Liu}, T. 2013, \apjs, 209, 37

\bibitem[{{Men'shchikov}(2016)}]{Menshchikov2016}
{Men'shchikov}, A. 2016, \aap, 593, A71

\bibitem[{{Molinari} {et~al.}(2010{\natexlab{a}}){Molinari}, {Swinyard},
  {Bally}, {Barlow}, {Bernard}, {Martin}, {Moore}, {Noriega-Crespo}, {Plume},
  {Testi}, {Zavagno}, {Abergel}, {Ali}, {Anderson}, {Andr{\'e}}, {Baluteau},
  {Battersby}, {Beltr{\'a}n}, {Benedettini}, {Billot}, {Blommaert}, {Bontemps},
  {Boulanger}, {Brand}, {Brunt}, {Burton}, {Calzoletti}, {Carey}, {Caselli},
  {Cesaroni}, {Cernicharo}, {Chakrabarti}, {Chrysostomou}, {Cohen},
  {Compiegne}, {de Bernardis}, {de Gasperis}, {di Giorgio}, {Elia}, {Faustini},
  {Flagey}, {Fukui}, {Fuller}, {Ganga}, {Garcia-Lario}, {Glenn}, {Goldsmith},
  {Griffin}, {Hoare}, {Huang}, {Ikhenaode}, {Joblin}, {Joncas}, {Juvela},
  {Kirk}, {Lagache}, {Li}, {Lim}, {Lord}, {Marengo}, {Marshall}, {Masi},
  {Massi}, {Matsuura}, {Minier}, {Miville-Desch{\^e}nes}, {Montier}, {Morgan},
  {Motte}, {Mottram}, {M{\"u}ller}, {Natoli}, {Neves}, {Olmi}, {Paladini},
  {Paradis}, {Parsons}, {Peretto}, {Pestalozzi}, {Pezzuto}, {Piacentini},
  {Piazzo}, {Polychroni}, {Pomar{\`e}s}, {Popescu}, {Reach}, {Ristorcelli},
  {Robitaille}, {Robitaille}, {Rod{\'o}n}, {Roy}, {Royer}, {Russeil},
  {Saraceno}, {Sauvage}, {Schilke}, {Schisano}, {Schneider}, {Schuller},
  {Schulz}, {Sibthorpe}, {Smith}, {Smith}, {Spinoglio}, {Stamatellos},
  {Strafella}, {Stringfellow}, {Sturm}, {Taylor}, {Thompson}, {Traficante},
  {Tuffs}, {Umana}, {Valenziano}, {Vavrek}, {Veneziani}, {Viti}, {Waelkens},
  {Ward-Thompson}, {White}, {Wilcock}, {Wyrowski}, {Yorke}, \&
  {Zhang}}]{Molinari2010}
{Molinari}, S., {Swinyard}, B., {Bally}, J., {et~al.} 2010{\natexlab{a}}, \aap,
  518, L100

\bibitem[{{Molinari} {et~al.}(2010{\natexlab{b}}){Molinari}, {Swinyard},
  {Bally}, {Barlow}, {Bernard}, {Martin}, {Moore}, {Noriega-Crespo}, {Plume},
  {Testi}, {Zavagno}, {Abergel}, {Ali}, {Anderson}, {Andr{\'e}}, {Baluteau},
  {Battersby}, {Beltr{\'a}n}, {Benedettini}, {Billot}, {Blommaert}, {Bontemps},
  {Boulanger}, {Brand}, {Brunt}, {Burton}, {Calzoletti}, {Carey}, {Caselli},
  {Cesaroni}, {Cernicharo}, {Chakrabarti}, {Chrysostomou}, {Cohen},
  {Compiegne}, {de Bernardis}, {de Gasperis}, {di Giorgio}, {Elia}, {Faustini},
  {Flagey}, {Fukui}, {Fuller}, {Ganga}, {Garcia-Lario}, {Glenn}, {Goldsmith},
  {Griffin}, {Hoare}, {Huang}, {Ikhenaode}, {Joblin}, {Joncas}, {Juvela},
  {Kirk}, {Lagache}, {Li}, {Lim}, {Lord}, {Marengo}, {Marshall}, {Masi},
  {Massi}, {Matsuura}, {Minier}, {Miville-Desch{\^e}nes}, {Montier}, {Morgan},
  {Motte}, {Mottram}, {M{\"u}ller}, {Natoli}, {Neves}, {Olmi}, {Paladini},
  {Paradis}, {Parsons}, {Peretto}, {Pestalozzi}, {Pezzuto}, {Piacentini},
  {Piazzo}, {Polychroni}, {Pomar{\`e}s}, {Popescu}, {Reach}, {Ristorcelli},
  {Robitaille}, {Robitaille}, {Rod{\'o}n}, {Roy}, {Royer}, {Russeil},
  {Saraceno}, {Sauvage}, {Schilke}, {Schisano}, {Schneider}, {Schuller},
  {Schulz}, {Sibthorpe}, {Smith}, {Smith}, {Spinoglio}, {Stamatellos},
  {Strafella}, {Stringfellow}, {Sturm}, {Taylor}, {Thompson}, {Traficante},
  {Tuffs}, {Umana}, {Valenziano}, {Vavrek}, {Veneziani}, {Viti}, {Waelkens},
  {Ward-Thompson}, {White}, {Wilcock}, {Wyrowski}, {Yorke}, \&
  {Zhang}}]{Molinari_higal}
{Molinari}, S., {Swinyard}, B., {Bally}, J., {et~al.} 2010{\natexlab{b}}, \aap,
  518, L100

\bibitem[{{Myers}(2009)}]{Myers2009}
{Myers}, P.~C. 2009, \apj, 706, 1341

\bibitem[{{Myers}(2011)}]{Myers2011}
{Myers}, P.~C. 2011, \apj, 735, 82

\bibitem[{{Oya} {et~al.}(2014){Oya}, {Sakai}, {Sakai}, {Watanabe}, {Hirota},
  {Lindberg}, {Bisschop}, {J{\o}rgensen}, {van Dishoeck}, \&
  {Yamamoto}}]{Oya2014}
{Oya}, Y., {Sakai}, N., {Sakai}, T., {et~al.} 2014, \apj, 795, 152

\bibitem[{{Padoan} {et~al.}(2001){Padoan}, {Juvela}, {Goodman}, \&
  {Nordlund}}]{Padoan2001}
{Padoan}, P., {Juvela}, M., {Goodman}, A.~A., \& {Nordlund}, {\AA}. 2001, \apj,
  553, 227

\bibitem[{{Peretto} {et~al.}(2013){Peretto}, {Fuller}, {Duarte-Cabral},
  {Avison}, {Hennebelle}, {Pineda}, {Andr{\'e}}, {Bontemps}, {Motte},
  {Schneider}, \& {Molinari}}]{Peretto2013}
{Peretto}, N., {Fuller}, G.~A., {Duarte-Cabral}, A., {et~al.} 2013, \aap, 555,
  A112

\bibitem[{{Planck Collaboration}(2011{\natexlab{a}})}]{PlanckXXII}
{Planck Collaboration}. 2011{\natexlab{a}}, \aap, 536, A22

\bibitem[{{Planck Collaboration}(2011{\natexlab{b}})}]{PlanckXXIII}
{Planck Collaboration}. 2011{\natexlab{b}}, \aap, 536, A23

\bibitem[{{Planck Collaboration}(2011{\natexlab{c}})}]{PlanckXXV}
{Planck Collaboration}. 2011{\natexlab{c}}, \aap, 536, A25

\bibitem[{{Planck Collaboration} {et~al.}(2016){Planck Collaboration}, {Ade},
  {Aghanim}, {Arnaud}, {Ashdown}, {Aumont}, {Baccigalupi}, {Banday},
  {Barreiro}, {Bartolo}, {Battaner}, {Benabed}, {Beno{\^\i}t},
  {Benoit-L{\'e}vy}, {Bernard}, {Bersanelli}, {Bielewicz}, {Bonaldi},
  {Bonavera}, {Bond}, {Borrill}, {Bouchet}, {Boulanger}, {Bucher}, {Burigana},
  {Butler}, {Calabrese}, {Catalano}, {Chamballu}, {Chiang}, {Christensen},
  {Clements}, {Colombi}, {Colombo}, {Combet}, {Couchot}, {Coulais}, {Crill},
  {Curto}, {Cuttaia}, {Danese}, {Davies}, {Davis}, {de Bernardis}, {de Rosa},
  {de Zotti}, {Delabrouille}, {D{\'e}sert}, {Dickinson}, {Diego}, {Dole},
  {Donzelli}, {Dor{\'e}}, {Douspis}, {Ducout}, {Dupac}, {Efstathiou}, {Elsner},
  {En{\ss}lin}, {Eriksen}, {Falgarone}, {Fergusson}, {Finelli}, {Forni},
  {Frailis}, {Fraisse}, {Franceschi}, {Frejsel}, {Galeotta}, {Galli}, {Ganga},
  {Giard}, {Giraud-H{\'e}raud}, {Gjerl{\o}w}, {Gonz{\'a}lez-Nuevo},
  {G{\'o}rski}, {Gratton}, {Gregorio}, {Gruppuso}, {Gudmundsson}, {Hansen},
  {Hanson}, {Harrison}, {Helou}, {Henrot-Versill{\'e}},
  {Hern{\'a}ndez-Monteagudo}, {Herranz}, {Hildebrandt}, {Hivon}, {Hobson},
  {Holmes}, {Hornstrup}, {Hovest}, {Huffenberger}, {Hurier}, {Jaffe}, {Jaffe},
  {Jones}, {Juvela}, {Keih{\"a}nen}, {Keskitalo}, {Kisner}, {Knoche}, {Kunz},
  {Kurki-Suonio}, {Lagache}, {Lamarre}, {Lasenby}, {Lattanzi}, {Lawrence},
  {Leonardi}, {Lesgourgues}, {Levrier}, {Liguori}, {Lilje}, {Linden-V{\o}rnle},
  {L{\'o}pez-Caniego}, {Lubin}, {Mac{\'\i}as-P{\'e}rez}, {Maggio}, {Maino},
  {Mand olesi}, {Mangilli}, {Marshall}, {Martin}, {Mart{\'\i}nez-Gonz{\'a}lez},
  {Masi}, {Matarrese}, {Mazzotta}, {McGehee}, {Melchiorri}, {Mendes},
  {Mennella}, {Migliaccio}, {Mitra}, {Miville-Desch{\^e}nes}, {Moneti},
  {Montier}, {Morgante}, {Mortlock}, {Moss}, {Munshi}, {Murphy}, {Naselsky},
  {Nati}, {Natoli}, {Netterfield}, {N{\o}rgaard-Nielsen}, {Noviello},
  {Novikov}, {Novikov}, {Oxborrow}, {Paci}, {Pagano}, {Pajot}, {Paladini},
  {Paoletti}, {Pasian}, {Patanchon}, {Pearson}, {Pelkonen}, {Perdereau},
  {Perotto}, {Perrotta}, {Pettorino}, {Piacentini}, {Piat}, {Pierpaoli},
  {Pietrobon}, {Plaszczynski}, {Pointecouteau}, {Polenta}, {Pratt},
  {Pr{\'e}zeau}, {Prunet}, {Puget}, {Rachen}, {Reach}, {Rebolo}, {Reinecke},
  {Remazeilles}, {Renault}, {Renzi}, {Ristorcelli}, {Rocha}, {Rosset},
  {Rossetti}, {Roudier}, {Rubi{\~n}o-Mart{\'\i}n}, {Rusholme}, {Sandri},
  {Santos}, {Savelainen}, {Savini}, {Scott}, {Seiffert}, {Shellard}, {Spencer},
  {Stolyarov}, {Sudiwala}, {Sunyaev}, {Sutton}, {Suur-Uski}, {Sygnet},
  {Tauber}, {Terenzi}, {Toffolatti}, {Tomasi}, {Tristram}, {Tucci}, {Tuovinen},
  {Umana}, {Valenziano}, {Valiviita}, {Van Tent}, {Vielva}, {Villa}, {Wade},
  {Wandelt}, {Wehus}, {Yvon}, {Zacchei}, \& {Zonca}}]{PGCC2016}
{Planck Collaboration}, {Ade}, P.~A.~R., {Aghanim}, N., {et~al.} 2016, \aap,
  594, A28

\bibitem[{{Poglitsch} {et~al.}(2010){Poglitsch}, {Waelkens}, {Geis},
  {Feuchtgruber}, {Vandenbussche}, {Rodriguez}, {Krause}, {Renotte}, {van
  Hoof}, {Saraceno}, {Cepa}, {Kerschbaum}, {Agn{\`e}se}, {Ali}, {Altieri},
  {Andreani}, {Augueres}, {Balog}, {Barl}, {Bauer}, {Belbachir}, {Benedettini},
  {Billot}, {Boulade}, {Bischof}, {Blommaert}, {Callut}, {Cara}, {Cerulli},
  {Cesarsky}, {Contursi}, {Creten}, {De Meester}, {Doublier}, {Doumayrou},
  {Duband}, {Exter}, {Genzel}, {Gillis}, {Gr{\"o}zinger}, {Henning},
  {Herreros}, {Huygen}, {Inguscio}, {Jakob}, {Jamar}, {Jean}, {de Jong},
  {Katterloher}, {Kiss}, {Klaas}, {Lemke}, {Lutz}, {Madden}, {Marquet},
  {Martignac}, {Mazy}, {Merken}, {Montfort}, {Morbidelli}, {M{\"u}ller},
  {Nielbock}, {Okumura}, {Orfei}, {Ottensamer}, {Pezzuto}, {Popesso},
  {Putzeys}, {Regibo}, {Reveret}, {Royer}, {Sauvage}, {Schreiber}, {Stegmaier},
  {Schmitt}, {Schubert}, {Sturm}, {Thiel}, {Tofani}, {Vavrek}, {Wetzstein},
  {Wieprecht}, \& {Wiezorrek}}]{Poglitsch2010}
{Poglitsch}, A., {Waelkens}, C., {Geis}, N., {et~al.} 2010, \aap, 518, L2

\bibitem[{{Reid} {et~al.}(2016){Reid}, {Dame}, {Menten}, \&
  {Brunthaler}}]{Reid2016Bayesian}
{Reid}, M.~J., {Dame}, T.~M., {Menten}, K.~M., \& {Brunthaler}, A. 2016, \apj,
  823, 77

\bibitem[{{Sadavoy} {et~al.}(2013){Sadavoy}, {Di Francesco}, {Johnstone},
  {Currie}, {Drabek}, {Hatchell}, {Nutter}, {Andr{\'e}}, {Arzoumanian},
  {Benedettini}, {Bernard}, {Duarte-Cabral}, {Fallscheer}, {Friesen},
  {Greaves}, {Hennemann}, {Hill}, {Jenness}, {K{\"o}nyves}, {Matthews},
  {Mottram}, {Pezzuto}, {Roy}, {Rygl}, {Schneider-Bontemps}, {Spinoglio},
  {Testi}, {Tothill}, {Ward-Thompson}, {White}, {JCMT}, \& {Herschel Gould Belt
  Survey Teams}}]{Sadavoy2013}
{Sadavoy}, S.~I., {Di Francesco}, J., {Johnstone}, D., {et~al.} 2013, \apj,
  767, 126

\bibitem[{{Shan} {et~al.}(2012){Shan}, {Yang}, {Shi}, {Yao}, {Zuo}, {Lin},
  {Chen}, {Zhang}, {Duan}, {Cao}, {Li}, {Li}, {Liu}, \& {Zhong}}]{Shan2012}
{Shan}, W., {Yang}, J., {Shi}, S., {et~al.} 2012, IEEE Transactions on
  Terahertz Science and Technology, 2, 593

\bibitem[{{Shinnaga} {et~al.}(2004){Shinnaga}, {Ohashi}, {Lee}, \&
  {Moriarty-Schieven}}]{Shinnaga2004}
{Shinnaga}, H., {Ohashi}, N., {Lee}, S.-W., \& {Moriarty-Schieven}, G.~H. 2004,
  \apj, 601, 962

\bibitem[{{Tan} {et~al.}(2014){Tan}, {Beltr{\'a}n}, {Caselli}, {Fontani},
  {Fuente}, {Krumholz}, {McKee}, \& {Stolte}}]{Tan_PPVI2014}
{Tan}, J.~C., {Beltr{\'a}n}, M.~T., {Caselli}, P., {et~al.} 2014, Protostars
  and Planets VI, 149

\bibitem[{{Wang} {et~al.}(2018){Wang}, {Zahorecz}, {Cunningham}, {T{\'o}th},
  {Liu}, {Lu}, {Wang}, {Cosentino}, {Sung}, {Sokolov}, {Wang}, {Wang}, {Zhang},
  {Li}, {Kim}, {Tatematsu}, {Testi}, {Wu}, {Yang}, \& {SAMPLING
  Collaboration}}]{Wang2018}
{Wang}, K., {Zahorecz}, S., {Cunningham}, M.~R., {et~al.} 2018, Research Notes
  of the American Astronomical Society, 2, 2

\bibitem[{{Wang} {et~al.}(2010){Wang}, {Li}, {Abel}, \& {Nakamura}}]{Wang2010}
{Wang}, P., {Li}, Z.-Y., {Abel}, T., \& {Nakamura}, F. 2010, \apj, 709, 27

\bibitem[{{Ward-Thompson} {et~al.}(2016){Ward-Thompson}, {Pattle}, {Kirk},
  {Marsh}, {Buckle}, {Hatchell}, {Nutter}, {Griffin}, {Di Francesco},
  {Andr{\'e}}, {Beaulieu}, {Berry}, {Broekhoven-Fiene}, {Currie}, {Fich},
  {Jenness}, {Johnstone}, {Kirk}, {Mottram}, {Pineda}, {Quinn}, {Sadavoy},
  {Salji}, {Tisi}, {Walker-Smith}, {White}, {Hill}, {K{\"o}nyves}, {Palmeirim},
  \& {Pezzuto}}]{Ward-Thompson2016}
{Ward-Thompson}, D., {Pattle}, K., {Kirk}, J.~M., {et~al.} 2016, \mnras, 463,
  1008

\bibitem[{{Whitworth} \& {Ward-Thompson}(2001)}]{Whitworth2001}
{Whitworth}, A.~P. \& {Ward-Thompson}, D. 2001, \apj, 547, 317

\bibitem[{{Wu} {et~al.}(2012){Wu}, {Liu}, {Meng}, {Li}, {Qin}, \&
  {Ju}}]{Wu2012}
{Wu}, Y., {Liu}, T., {Meng}, F., {et~al.} 2012, \apj, 756, 76

\bibitem[{{Yuan} {et~al.}(2018){Yuan}, {Li}, {Wu}, {Ellingsen}, {Henkel},
  {Wang}, {Liu}, {Liu}, {Zavagno}, {Ren}, \& {Huang}}]{Yuan2018}
{Yuan}, J., {Li}, J.-Z., {Wu}, Y., {et~al.} 2018, \apj, 852, 12

\end{thebibliography}

\begin{appendix}

\section{Additional figures}

\begin{figure*}
\sidecaption
\includegraphics[width=11.5cm]{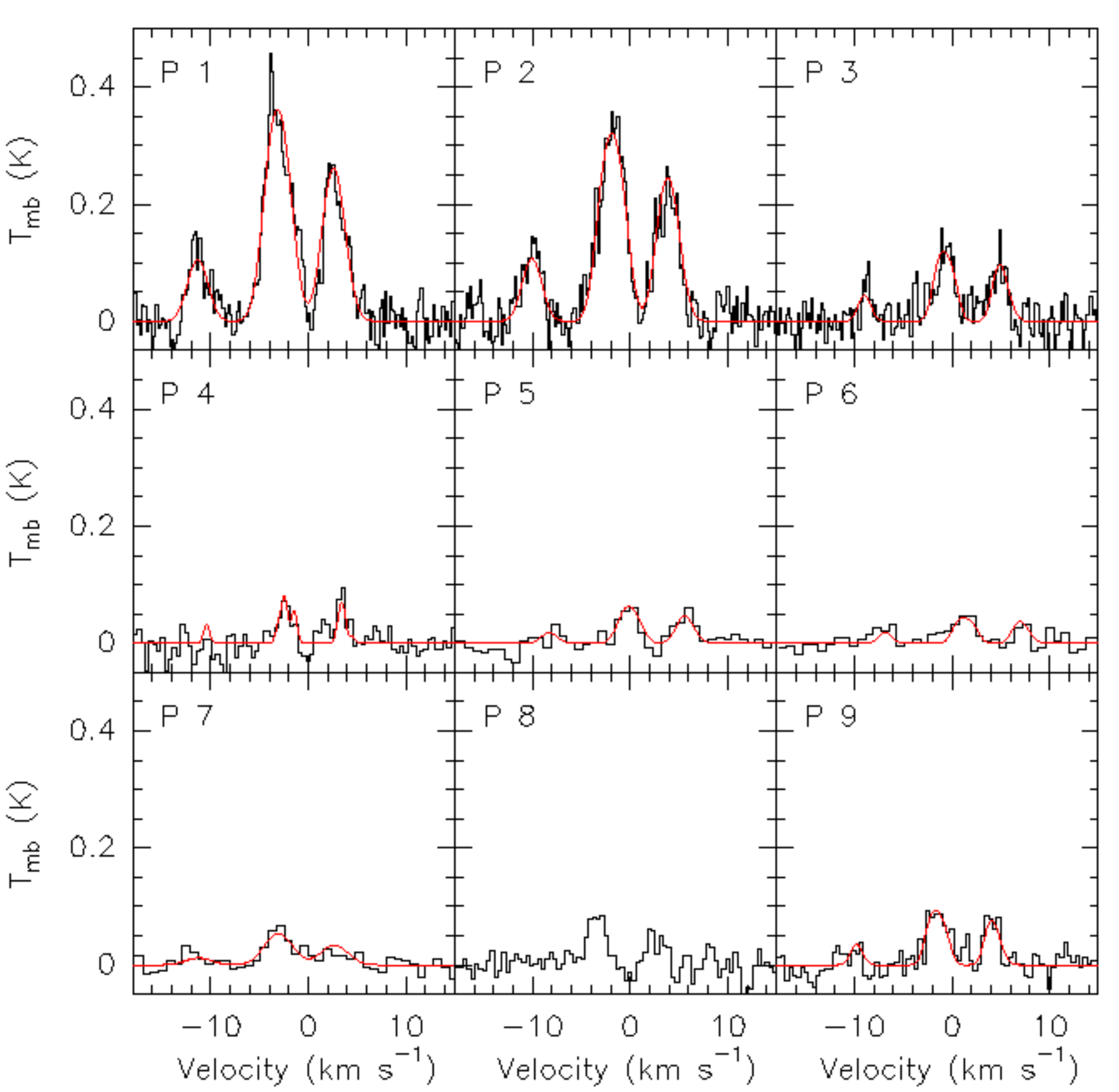}
\caption{$\rm N_2H^{+}$ spectra towards the nine pointings shown in Fig. \ref{fig:KVN_points}. The red curves show fits to the hyperfine structure of the line components. For pointing 8, the fit fails to converge.}
\label{fig:KVN_n2hp}
\end{figure*}

\begin{figure*}
\sidecaption
\includegraphics[width=11.5cm]{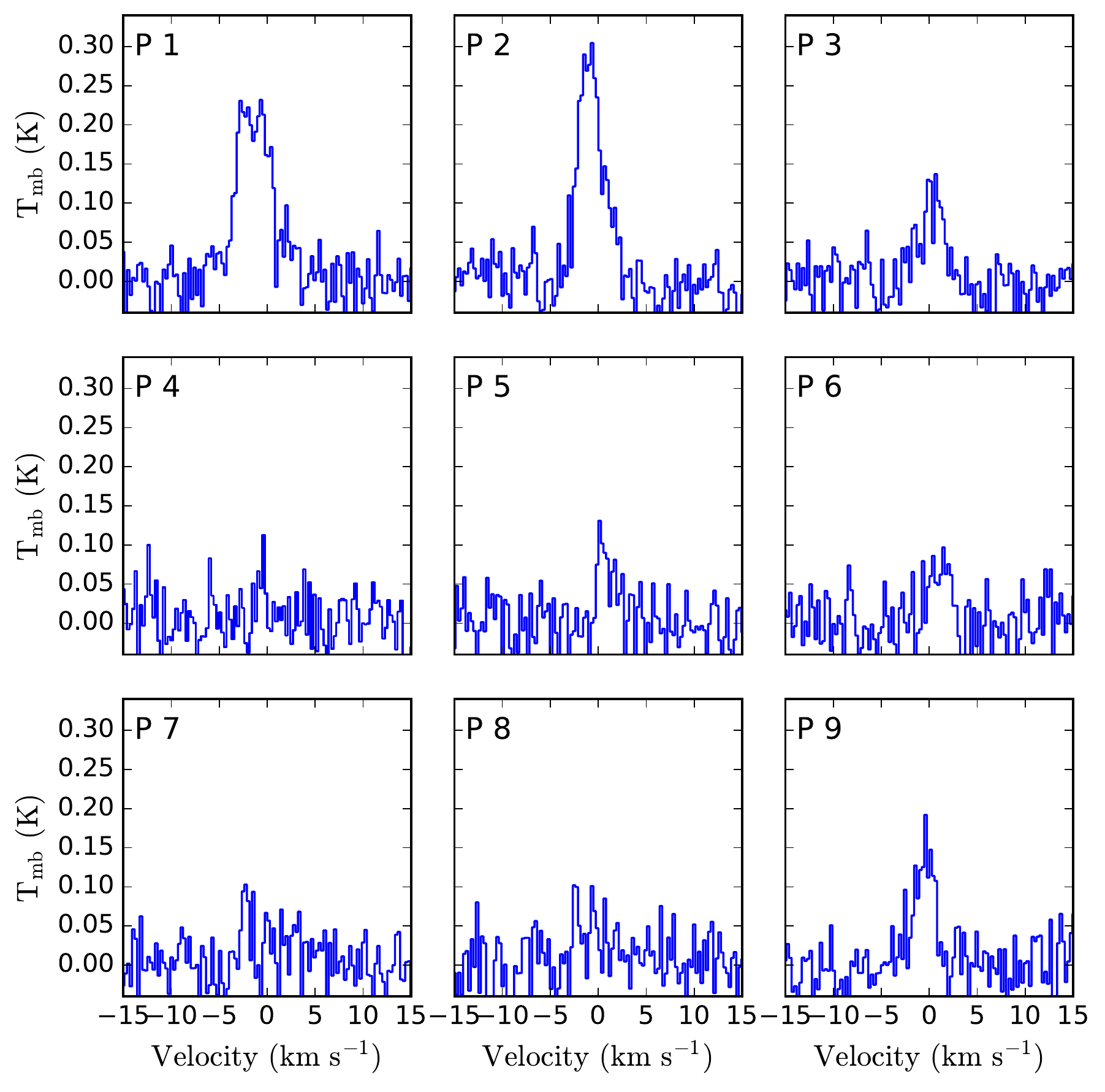}
\caption{$\rm H_2CO$ spectra towards the nine pointings shown in Fig. \ref{fig:KVN_points}.}
\label{fig:KVN_h2co}
\end{figure*}

\begin{figure*}
\includegraphics[width=8.5cm]{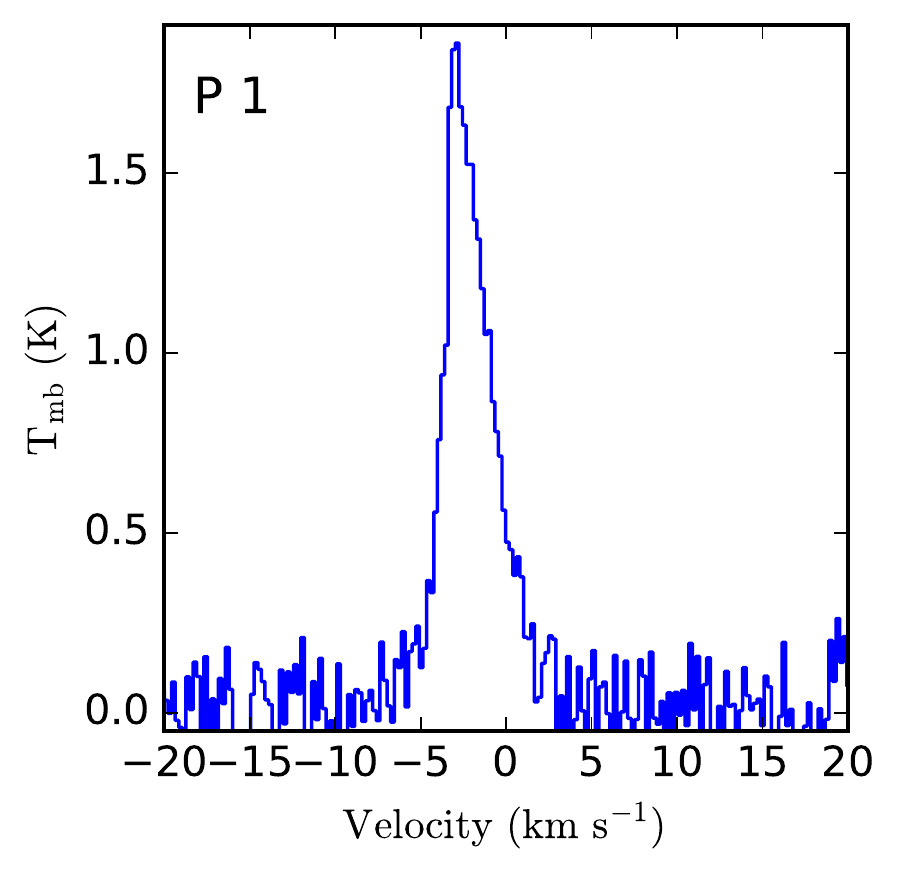}
\caption{$\rm HCO^{+}$ spectra towards pointing 1 in Fig. \ref{fig:KVN_points}.}
\label{fig:KVN_hco}
\end{figure*}

Figures \ref{fig:KVN_n2hp} and \ref{fig:KVN_h2co} show the $\rm N_2H^{+}$ and $\rm H_2CO$ spectra extracted from the nine pointings shown in Fig. \ref{fig:KVN_points} using the KVN antenna. Only the spectra from pointings 1 to 4 is used in our analysis as the S/N ratio of the pointings 5-9 is low. Shown in Fig. \ref{fig:KVN_hco} is the single $\rm HCO^{+}$ spectra extracted from pointing 1.

\begin{figure*}
\sidecaption
\includegraphics[width=12cm]{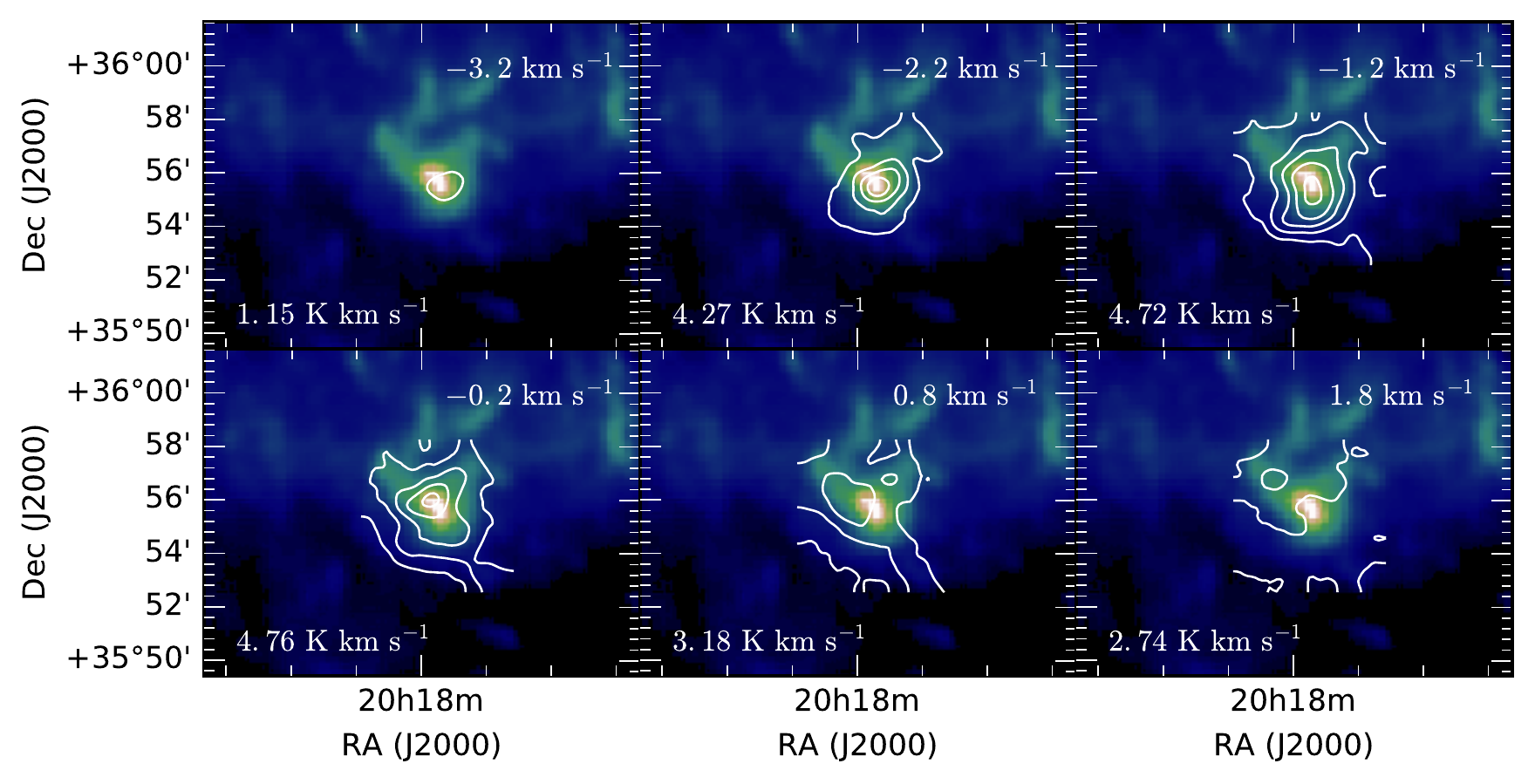}
\caption{Channel maps of the $^{13}\rm CO$ $J=2-1$ line emission superposed on the $N(\rm H_2)$ map. The line observations, white contours, are from the PMO and are integrated over a 1.0 km\,s$^{-1}$ wide velocity intervals. The velocity centroids are indicated in the top righ and the maximum integrated intensity in the channel is indicated in the bottom left.}
\label{fig:PMO_13CO}
\end{figure*}

Shown in Fig. \ref{fig:PMO_13CO} are the PMO $^{13}\rm CO$ $J=2-1$ line observations, integrated over 1.0 km\,s$^{-1}$ wide velocity intervals. The underlying colour map is the $\rm H_2$ column density map, derived from the \textit{Herschel}/SPIRE observations.

\begin{figure*}
\includegraphics[width=17.8cm]{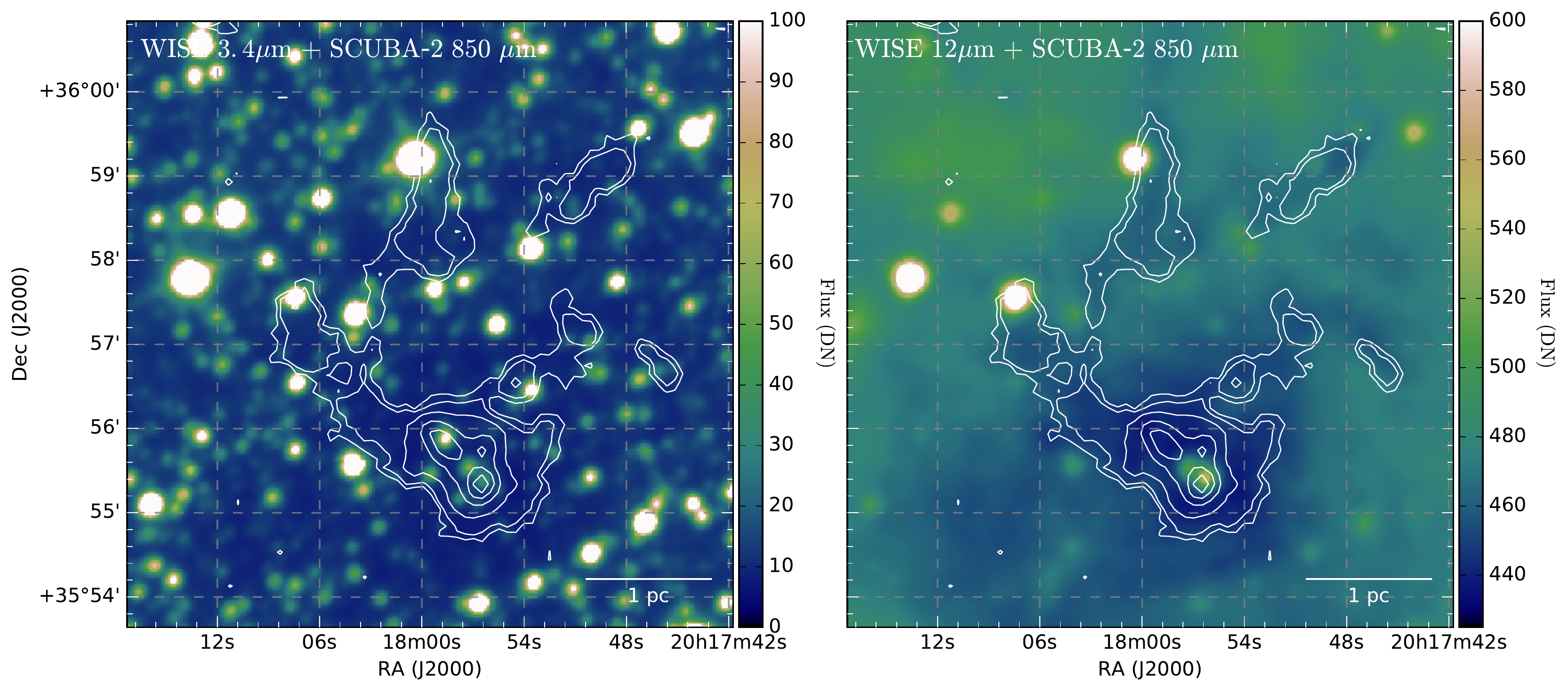}
\caption{3.4 $\mu$m (left panel) and 12 $\mu$m (right panel) images of G074.11+00.11 obtained with the WISE satellite. The white contours show the SCUBA-2 map at 850 $\mu$m. The WISE observations are in the native Digital Numbers (DN).}
\label{fig:WISE}
\end{figure*}

Shown in Fig. \ref{fig:WISE} are the 3.4 and 12 $\mu$m observations from the WISE satellite. The central clump is associated with several NIR/MIR point sources and is seen in absorption at 12 $\mu$m. The SCUBA-2 observations trace extremely well the regions seen in absorption.

\section{Distance estimation}

To estimate the mass of the region, a reliable distance estimate is needed. The distance reported in the PGCC catalogue is 3.45 kpc \citep{PGCC2016}. This is derived from NIR extinction. On the other hand, assuming a $V_{\rm LSR} = -2$ $\rm km\,s^{-1}$ for the clump, a Bayesian distance estimate \citep{Reid2016Bayesian} with no prior information whether the source is located near or far, $P_{\rm far} = 0.5$, gives the highest probability ($P=0.9$) for a distance of 4.3 $\pm$ 0.7 kpc. The second and third highest probabilities given by the Bayesian method are 2.4 and 1.3 kpc, both of which have a similar probability ($P=0.05$ and $P=0.03$, respectively). However, estimating distances from kinematics can be unreliable in the Galactic Plane, thus, we derive an independent estimate for the distance using a three dimensional extinction method \citep{Marshall2006}. The best fit from the method gives two components in the line-of-sight, one at 0.96 $\pm$ 0.07 kpc and one at 2.3 $\pm$ 0.1 kpc. The estimated visual extinction values given by the method for these components are $A_v =$ 1.7 magnitudes and $A_v$ $\sim 10$ magnitudes, respectively. 

\begin{figure*}
\sidecaption
\includegraphics[width=11.5cm]{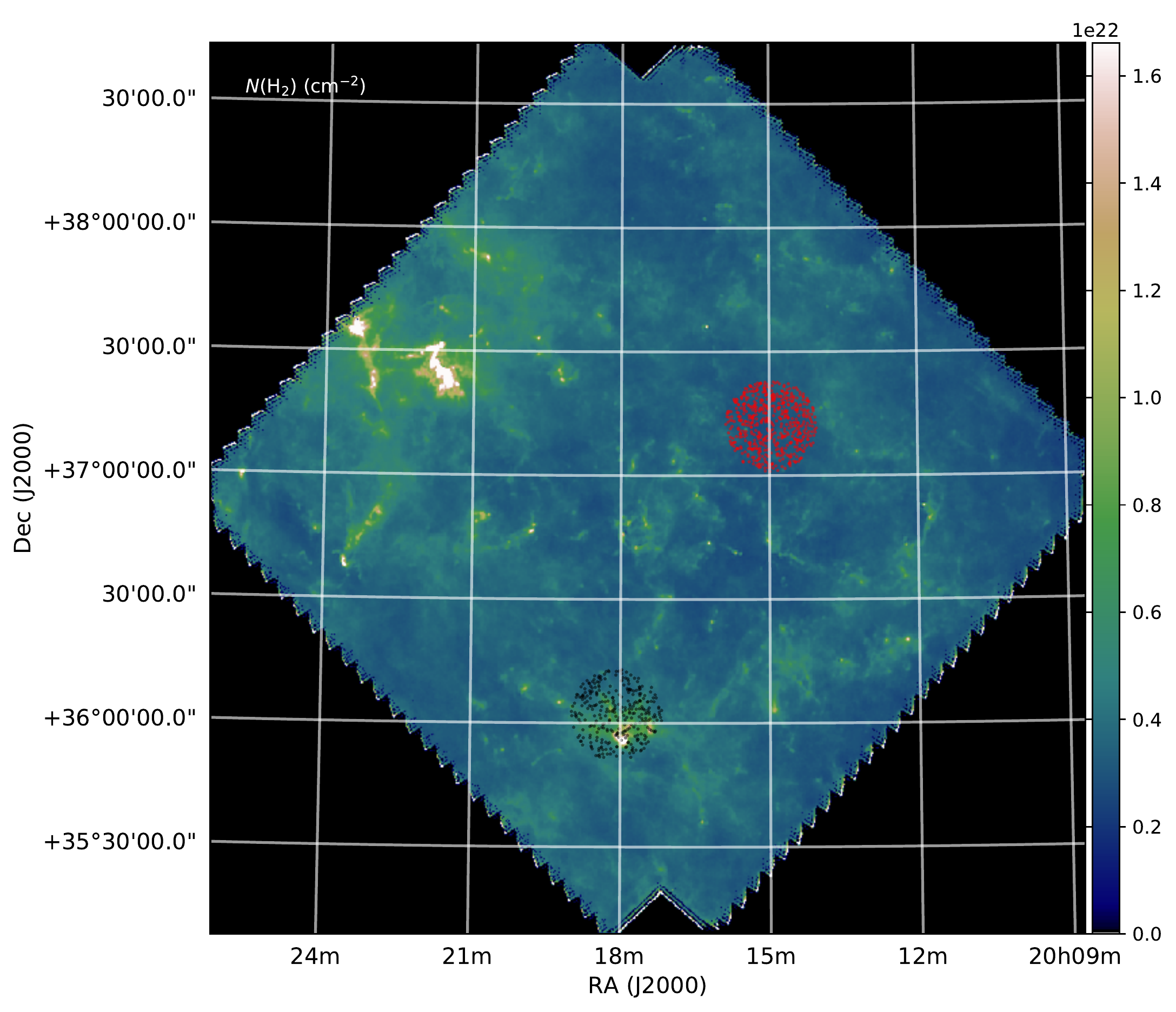}
\caption{Locations of the stars from the Gaia DR2 catalogue used in our distance estimation. The stars in our "Cloud" position are indicated with black markers, and those in the "Reference" position are indicated with red.}
\label{fig:gaia_locs}
\end{figure*}

Since the above methods give three possible distances, although the probability from the Bayesian estimate for the distance of $\sim$ 2 kpc is low, we derive a third independent estimate from the Gaia observations \citep{GAIA2016}. The Gaia data release 2 \citep[DR2][]{GAIA_DR2_2018} gives the positions, parallaxes, and magnitudes for more than 1.6 billion sources, and provides an excellent tool for distance estimation. 

\begin{figure*}
\sidecaption
\includegraphics[width=11.5cm]{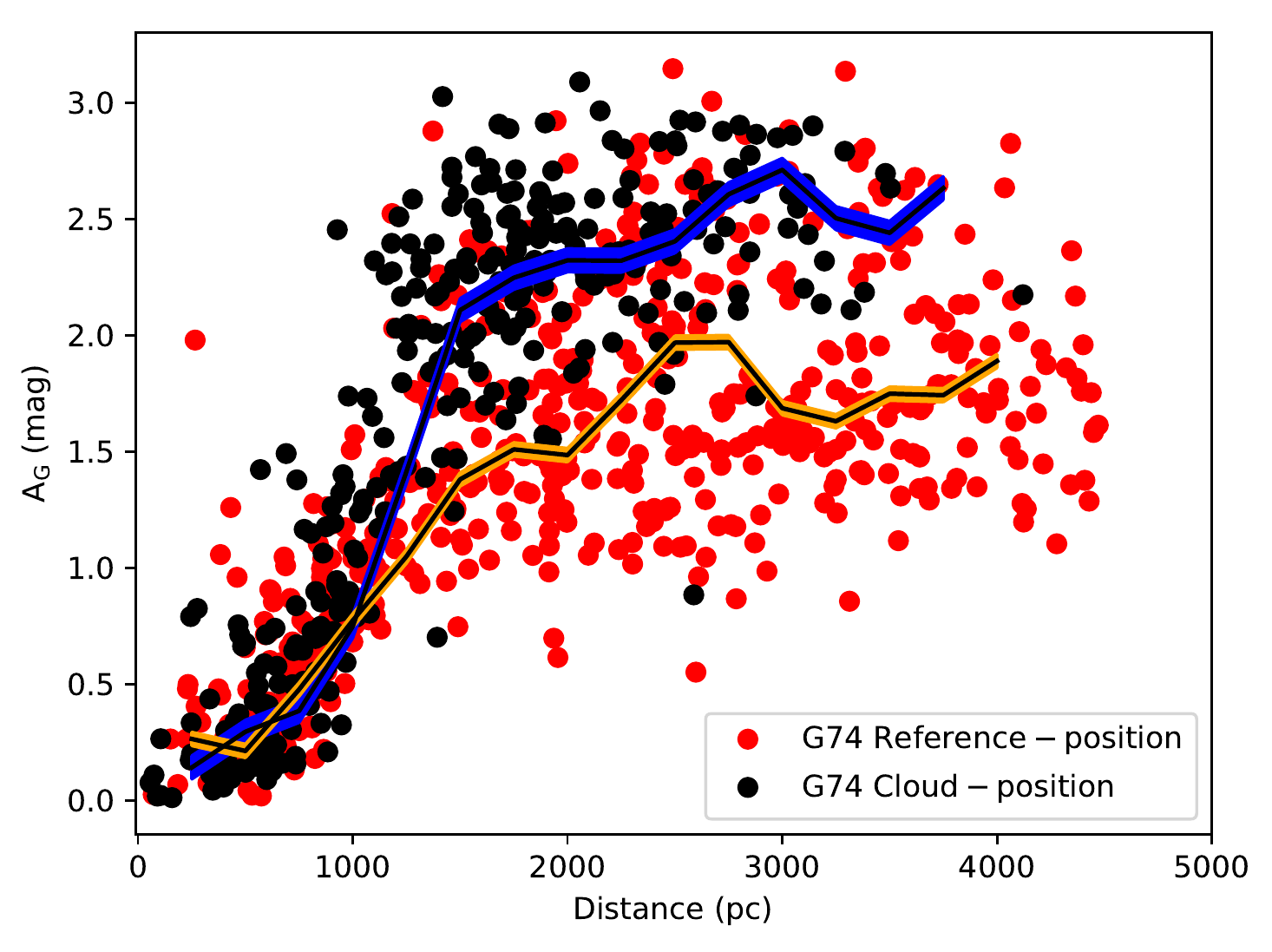}
\caption{Line-of-sight extinction in the Gaia G-band as a function of distance. The black and red points correspond to the markers in Fig. \ref{fig:gaia_locs}. The black-blue and black-orange lines show the median value of $A_{\rm G}$, computed over a 250 pc wide bins. The width of the blue and orange area corresponds to the error of the mean value of the data points within each bin.}
\label{fig:gaia_Ag}
\end{figure*}

Shown in Fig. \ref{fig:gaia_locs} are the locations of $\sim$1000 stars that are used in our distance estimation. The stars were selected from the DR2 catalogue based on two criteria: the value line-of-sight extinction in G band is not negative $A_{\rm G} > 0$, and the parallax-over-error is larger than five $\bar{\omega} / \sigma_{\bar{\omega}}  > 5$. The parallaxes were then converted to distances by $r=1/\bar{\omega}$, where $\bar{\omega}$ is given in seconds of arc. However, as discussed by \citet{Bailer-Jones2015, Luri2018}, a priori assumptions or Bayesian methods should be used when deriving distance estimates from the parallax measurements to get reliable distance estimates, but this exercise is beyond the scope of this paper. Furthermore, the $A_{\rm G}$ values reported in the DR2 catalogue are likely incorrect on the star-by-star level, but they should be usable for larger scale studies \citep{Andrae2018}.

The resulting distance as a function of $A_{\rm G}$ are shown in Fig. \ref{fig:gaia_Ag} for two positions which we call 'Cloud' and 'Reference' and are indicated in Fig. \ref{fig:gaia_locs}. The Cloud position covers our target G074.11+00.11. For the Cloud position stars, the median value of $A_{\rm G}$ over a 250 pc wide bins (the black-blue line) is relatively constant until a distance of $\sim$1 kpc, after which the extinction rapidly increases by a factor of $\sim4$ to a median value of 2.3 mag at $\sim$2 kpc. For the Reference position stars (black-orange line), the increase of extinction is not as steep, reaching a median value of 1.5 mag at 2 kpc.

Thus, assuming that the increase in the $A_{\rm G}$ is caused only by increasing column density in the line-of-sight, the lower limit for the distance to the region G074.11+00.11 is $\sim 1.25-1.75$ kpc. However, taking into account that both the Cloud and Reference positions show an almost constant median value of $A_{\rm G}$ at a distance of $\sim$2 kpc and that the best fit from the extinction method is at $\sim$2 kpc, we assume a distance of 2.3 kpc for the cloud.

\end{appendix}
\end{document}